\documentclass[runningheads]{llncs}

\usepackage{eccv}

\usepackage{orcidlink}
\usepackage{eccvabbrv}

\usepackage{graphicx}
\usepackage{booktabs}
\usepackage{multirow}
\usepackage{makecell}
\usepackage{wrapfig}
\usepackage{capt-of}
\usepackage{soul}
\usepackage{bm}
\usepackage{pifont}
\usepackage{tabularx}
\usepackage{placeins}
\usepackage{float}

\usepackage[accsupp]{axessibility}  %
\providecommand{\Description}[1]{} %
\providecommand{\cmark}{\ding{51}}
\providecommand{\xmark}{\ding{55}}

\usepackage{hyperref}

\providecommand{\lblfig}[1]{\label{fig:#1}}
\providecommand{\lblsec}[1]{\label{sec:#1}}
\providecommand{\lbltbl}[1]{\label{tbl:#1}}
\providecommand{\reffig}[1]{Fig.~\ref{fig:#1}}
\providecommand{\refsec}[1]{Sec.~\ref{sec:#1}}
\providecommand{\reftbl}[1]{Table~\ref{tbl:#1}}
\providecommand{\refapp}[1]{Appendix~\ref{sec:#1}}
\providecommand{\myparagraph}[1]{\paragraph{\normalfont\textbf{#1}}}

\def \mywebsitelink{https://ruihangao.github.io/Text2TactileGraphics/}

\begin{document}

\title{Text-based Tactile Graphics Generation for \\ the Visually Impaired}

\author{Ruihan Gao\inst{1}$^{\star}$ \and
Joonghyuk Shin \inst{1,2}$^{\star}$ \and
Ava Pun\inst{1} \and
Jaesik Park\inst{2} \and \\
Wenzhen Yuan\inst{3} \and
Jun-Yan Zhu\inst{1}
}

\authorrunning{R.~Gao et al.}

\institute{
Carnegie Mellon University, Pittsburgh, PA, USA
\and Seoul National University, Seoul, Republic of Korea
\and University of Illinois Urbana-Champaign, Urbana, IL, USA
}

\maketitle
{\renewcommand{\thefootnote}{}\NoHyper\footnotetext{$^{\star}$Equal contribution.}\endNoHyper}

\begin{abstract}
Tactile graphics are a primary medium for blind and low-vision (BLV) individuals to access non-textual information. 
However, they are difficult to scale or personalize.
While recent generative models have revolutionized visual content creation, they are optimized for screen-based visual realism and fail to satisfy the haptic perceptual and physical fabrication constraints required for touch.
We present the first integrated generative system that produces fabrication-ready 2.5D tactile graphics directly from natural language prompts, jointly generating global base geometry, fine-grained tactile surface textures, and standard-compliant braille within a unified 3D-printable representation. 
Our approach introduces fabrication-aware techniques, including template-guided relief generation, a fast diffusion-based text-to-texture module for high-resolution tileable normal maps, and strict base flattening to ensure tactile readability and printability, while supporting both automatic generation and interactive texture control. 
Extensive evaluations, together with in-person user studies with BLV participants and blindfolded sighted participants using physically 3D-printed outputs, show that participants consistently prefer our results over baselines. 
By extending generative graphics beyond screens to touchable reliefs, our work broadens access to generative AI for the BLV community and beyond. The code and project webpage are available at \url{\mywebsitelink}.
\end{abstract}

\begin{figure*}[t]
    \centering
    \includegraphics[width=\textwidth]{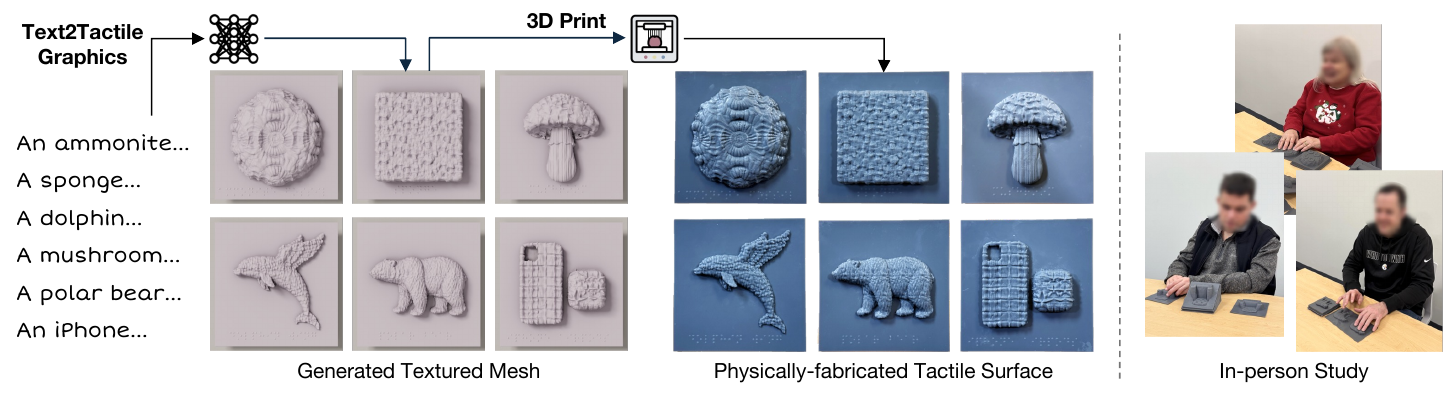}
    \caption{Given a text prompt describing an object and its surface textures, our method generates \emph{3D-printable 2.5D tactile reliefs} that integrate global base geometry, fine-grained surface textures, and readable braille annotations. 
    By jointly modeling geometry, texture, and braille within a unified fabrication-ready representation, our method produces perceptually salient tactile details and supports expressive, accessible exploration for blind and low-vision users. Please refer to our webpage for a video of an interactive demo and 3D visualizations.}
    \Description{A teaser figure depicts a set of generated tactile graphics, physical 3D-printed versions of the graphics, and several individuals exploring the graphics through touch.}
    \label{fig:teaser}
\end{figure*}

\section{Introduction}

Recent advances in generative models have dramatically expanded the scope of computer graphics, enabling text-driven synthesis of images~\cite{stable-diffusion-3,dall-e-2,imagen3}, videos~\cite{veo3,sora2}, and 3D assets~\cite{dreamfusion,magic3d,trellis2,hunyuan3d-2-5} with unprecedented realism and diversity. These systems have reshaped how visual content is created, edited,
and consumed, lowering barriers for artists, designers, and everyday users.

Despite this progress, modern generative pipelines remain almost exclusively \emph{visual}, designed for screen-based viewing. 
This overlooks a complementary and fundamentally physical modality: \emph{touch}. 
For blind and low-vision (BLV) individuals, who constitute over 290 million people worldwide, including 43 million who are fully blind~\cite{blindness-estimates}, touch is a primary means of understanding shapes, structures, and materials. 
Tactile graphics, which encode information through raised geometry, surface textures, and braille annotations, are widely used in education, accessibility, and cultural contexts such as diagrams, maps, and museum exhibits~\cite{gardner1996tactile,tactile-paintings,tactile-map-generation-review,tactile-textbooks}.

Despite their importance, tactile graphics are still largely produced through manual, expert-driven workflows such as swell paper, embossing, collage, and sculptural relief fabrication. 
Designers must interpret visual or textual content and translate it into touch-friendly representations, making the process time-consuming, labor-intensive, and difficult to scale. 
As a result, existing tactile graphic collections remain limited in scope and primarily focus on standardized educational materials~\cite{aph-library,perkins-library}. Compared to visual graphics, where AI enables rapid creation and personalization, tactile graphics offer little opportunity for user-driven or creative generation.

Tactile graphics also impose constraints not addressed by existing generative pipelines. 
Unlike images or meshes rendered for viewing, they must satisfy \emph{haptic perceptual constraints} such as tactile contrast, resolution, and style consistency~\cite{clepper2025would}, while remaining compatible with \emph{physical fabrication} processes like 3D printing. Simply generating a depth map or visually plausible mesh is insufficient. Current text-to-3D models often produce non-manifold surfaces, fragile structures, and disconnected geometry, while lacking fine tactile textures because they are optimized for global object shape rather than touch-perceptible detail. Consequently, they require extensive post-processing or fail fabrication entirely.

To address these challenges, we introduce a \emph{generative graphics system for tactile content creation}. 
Given text descriptions of the desired object and textures, our method generates a \emph{2.5D tactile graphic} integrating geometry, tactile texture, and braille annotations, directly ready for standard 3D printing. 
At its core is a text-to-tactile-texture module that produces fine-resolution surface geometry from text, targeting texture-centric generation rather than global shape synthesis; the outputs are converted into tileable normal and displacement maps for interactive application at arbitrary scales and regions. 
To ensure physical realizability, we design fabrication-aware processing that unifies template-guided base generation, flat-base enforcement, and standard-compliant braille placement, and provide an interactive interface supporting workflows from fully automatic generation to fine-grained control over texture type, scale, and spatial assignment. 
Experiments demonstrate that our system produces tactile graphics that are perceptually effective, fabrication-compatible, and faithful to input prompts, significantly lowering the barrier to personalized tactile content for the BLV community and more broadly opening new research directions at the intersection of generative modeling, accessibility, and physical fabrication.

\section{Related Work}
\lblsec{related_work}

\myparagraph{Tactile Graphics and Physical Accessibility Representations.}
Tactile graphics are raised-line drawings or textured representations that convey visual and spatial information through touch, enabling blind and low-vision (BLV) users to access diagrams, maps, and illustrations. Traditional tactile graphics are typically produced using swell paper, embossing, or sculptural bas-relief fabrication, but these methods require expert manual design and are time-consuming, limiting scalability and availability~\cite{clepper2025would,authority2011guidelines}. A substantial body of work has studied how tactile graphics should be designed to be interpretable through touch. Design guidelines emphasize abstraction, clear contour hierarchies, sufficient spacing, and the use of discriminable textures to avoid tactile clutter. Because touch is local and sequential, overly complex graphics can hinder comprehension. Prior studies highlight the importance of simplifying visual content rather than attempting a direct visual-to-tactile translation~\cite{gardner1996tactile}. Research has also explored how tactile graphics can convey 3D information using 2D or shallow-relief representations~\cite{panotopoulou2020tactile,tovar2025geometrical}. Together, this literature establishes tactile graphics as a distinct medium with unique perceptual and fabrication constraints.

\myparagraph{Visual Content Generation.} 
Generative models~\cite{goodfellow2014generative, kingma2013auto, karras2019style} have achieved remarkable success in synthesizing 2D and 3D content. 
While denoising diffusion models now dominate in text-to-image generation~\cite{ho2020denoising,stable-diffusion}, generative modeling has also expanded to 3D, evolving from early volumetric approaches~\cite{wu2016learning} to diverse representations such as points, meshes, and neural fields~\cite{diffusion-pcd,meshgpt,triplane-diffusion,dreamfusion,magic3d}. 
Large-scale 3D generative models have further improved visual quality and geometric detail, producing meshes with realistic appearance when rendered~\cite{trellis,trellis2,hunyuan3d,hunyuan3d-2,hunyuan3d-2-5}. 
Generative models have also been applied to text-to-texture generation~\cite{text2mat,xiong2025texgaussian,zhang2024dreammat,vecchio2024matfuse}.
Despite their success, these models are optimized mainly for visual realism and screen-based consumption. They do not enforce fabrication constraints such as watertightness, minimum feature thickness, or geometric simplicity, nor do they account for haptic perception. As a result, outputs may remain unsuitable for physical fabrication or tactile interpretation, motivating fabrication- and perception-aware generative pipelines.

\myparagraph{Generative and Computational Approaches for Tactile Content.}
A growing body of work explores computational methods for generating tactile content to support BLV users. Early approaches focused on algorithmic image-to-tactile translation using edge detection and thresholding techniques~\cite{automatic-img2tactile,krufka2006}. While simple to implement, these methods often generate cluttered or ambiguous tactile outputs that violate tactile design principles. To address these limitations, later systems incorporated higher-level image understanding. For example, Pic2Tac~\cite{pakenaite2024pic2tac}  uses object detection to replace foreground elements with readable tactile icons and fills the background with uniform textures.
More recent work applies image generation models directly to tactile graphics creation~\cite{dzhurynskyi2024artificial,khan2025tactilenet,genai-tactile}, showing the promise of generative AI. However, these methods still focus on two-dimensional embossed or swell graphics, which lack the expressivity and textural richness of 2.5D sculpture. Several works explore the automated production of 2.5D graphics from image, normal map, or 3D inputs through traditional~\cite{bas-relief-from-3d,bas-relief-from-normal,tvcg-bas-relief} or learning-based~\cite{monorelief,neural-portrait-bas-relief,reliefnet} methods. Nonetheless, these works focus on fine geometric detail that may not be readable by touch or resolvable by fabrication, leaving the resulting sculptures unsuitable for tactile graphics.
Recent works~\cite{gao2023controllable,gao2024tactile} leverage generative priors to produce 2D and 3D objects with aligned color and tactile textures, but they are limited to per-material optimization and fail to generate new material textures.
In contrast to prior work, we create 2.5D graphics suitable for touch-based interpretation and physical fabrication, generalizing to any new materials given a text prompt input.

\section{Method}
\begin{figure*}[ht]
    \centering
    \includegraphics[width=\textwidth]{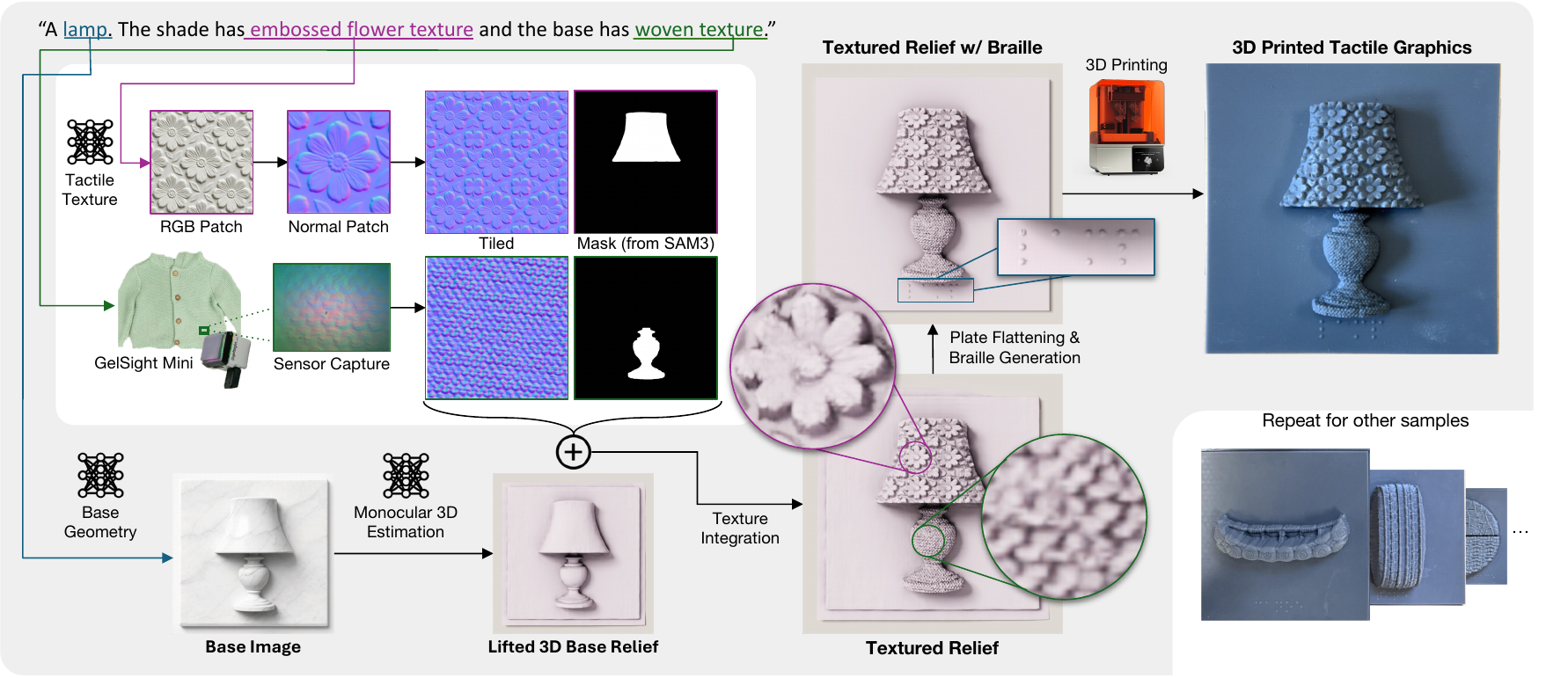}
    \caption{{\bf Method overview.} 
    Given a text prompt, we generate 3D-printable 2.5D tactile graphics that jointly encode geometry, surface texture, and braille.
    We first synthesize a base image constrained by a plate template and lift it to 3D base relief via monocular geometry estimation. 
    In parallel, a tactile texture module produces high-resolution, tileable normal maps from either text prompts or tactile sensor captures, which can be assigned to object parts through text or user interaction.
    The textures deform the base relief, followed by fabrication-aware base flattening and braille generation. The resulting watertight mesh is fabricated via SLA resin 3D printing.
    }
    \Description{Our pipeline involves base image generation, texture generation, lifting the base image to a 3D relief, adding textures and braille annotations, and finally physically fabricating the output through 3D printing.}
    \lblfig{method}
\end{figure*}

Our goal is to generate touch-friendly, fabrication-ready 2.5D tactile graphics from text prompts. This problem is challenging for three reasons:
(i) data scarcity, because no large-scale dataset models touch-suitable 2.5D graphics;
(ii) representation mismatch, since tactile graphics lie outside the distribution of object-centric 3D generation benchmarks, requiring both global shape abstraction and fine-scale surface relief suitable for touch; and
(iii) fabrication constraints, as existing methods do not explicitly enforce watertightness, flat bases, or geometric regularity required for haptic readability and physical printing.

To address these challenges, we propose a text-driven pipeline that generates a 3D-printable tactile graphic integrating three complementary components:
\textit{global geometry} capturing object shape, \textit{tactile surface textures} encoding fine-grained material cues, and \textit{braille annotations} providing semantic grounding. 
As shown in \reffig{method}, our pipeline begins by generating a 2.5D base geometry from text (\refsec{base-geometry}), synthesizes and applies tactile surface textures with flexible spatial control (\refsec{texture-gen}, \refsec{tiling}), enforces fabrication-aware geometric constraints and integrates braille annotations (\refsec{braille}), and finally outputs a watertight mesh suitable for interactive use and 3D printing (\refsec{interactive-control}).

\subsection{Base Geometry Generation}
\lblsec{base-geometry}
To generate the global shape of the tactile graphic, we first synthesize a text-conditioned image $I$ of a 2.5D relief sculpture using a fixed base-plate template $I_{\mathrm{bp}}$. By reformulating pure text-to-image generation as an in-context editing task conditioned on this template, we avoid the ambiguous scale and depth often produced by general-purpose models in object- or scene-centric compositions. 
As shown in \reffig{baseplate_texture} (Qwen-Image-Edit, 4-step), the template encourages the object to rest on a planar support, yielding a consistent spatial layout and enabling more stable depth estimation across diverse prompts. Similar trends are observed across other base models, including Qwen-Image-Edit (40-step, \reffig{t2i_base_plate_qwen40}) and Nano Banana Pro (\reffig{t2i_base_plate_nb}).
Given the generated image $I \in \mathbb{R}^{H \times W \times 3}$, we estimate a monocular depth map $D \in \mathbb{R}^{H \times W}$ using an off-the-shelf depth prediction model $f_{\text{depth}}$, then convert it into a 2.5D relief by mapping depth values to height displacements over the base plate.
This representation preserves salient object contours while remaining shallow enough for tactile exploration and efficient fabrication, provideing a stable geometric foundation for subsequent tactile texture refinement.

\subsection{Tactile Texture Synthesis}
\lblsec{texture-gen}

\myparagraph{Texture generation from text prompt.} A straightforward approach to incorporating surface textures is to generate an image from a single text prompt describing both geometry and material appearance, and then lift the image to 2.5D. In practice, this strategy leads to three major issues:
(i) text-to-image models prioritize global coherence over fine-grained detail, and text prompts lack the spatial specificity needed for precise texture placement;
(ii) when the desired texture is semantically inconsistent with the object (\textit{e.g.,} avocado skin on a dolphin), models often suppress the specified texture;
(iii) monocular depth estimation tends to smooth high-frequency surface details, as these models are optimized for overall geometry rather than subtle relief textures.

To address these limitations, we introduce a dedicated \emph{text-to-tactile-texture} module that synthesizes high-resolution, texture-centric surface representations from text prompts.
Unlike object-centric image generation, our goal is to model local, repeatable surface patterns suitable for tactile perception and geometric displacement. Since most text-to-image models are trained on object- and scene-level imagery, they are poorly suited for texture-focused synthesis. We therefore adapt an open-source image generation model~\cite{qwen-image} by fine-tuning it on a curated texture dataset emphasizing local surface detail. The dataset consists of real captured textures and synthetically generated samples aligned with our target distribution (\reffig{data_tiling}). 
Additional details on dataset construction and training are provided in \refapp{supple_dataset_training_details}.

\begin{figure*}[t]
    \centering
    \includegraphics[width=\textwidth]{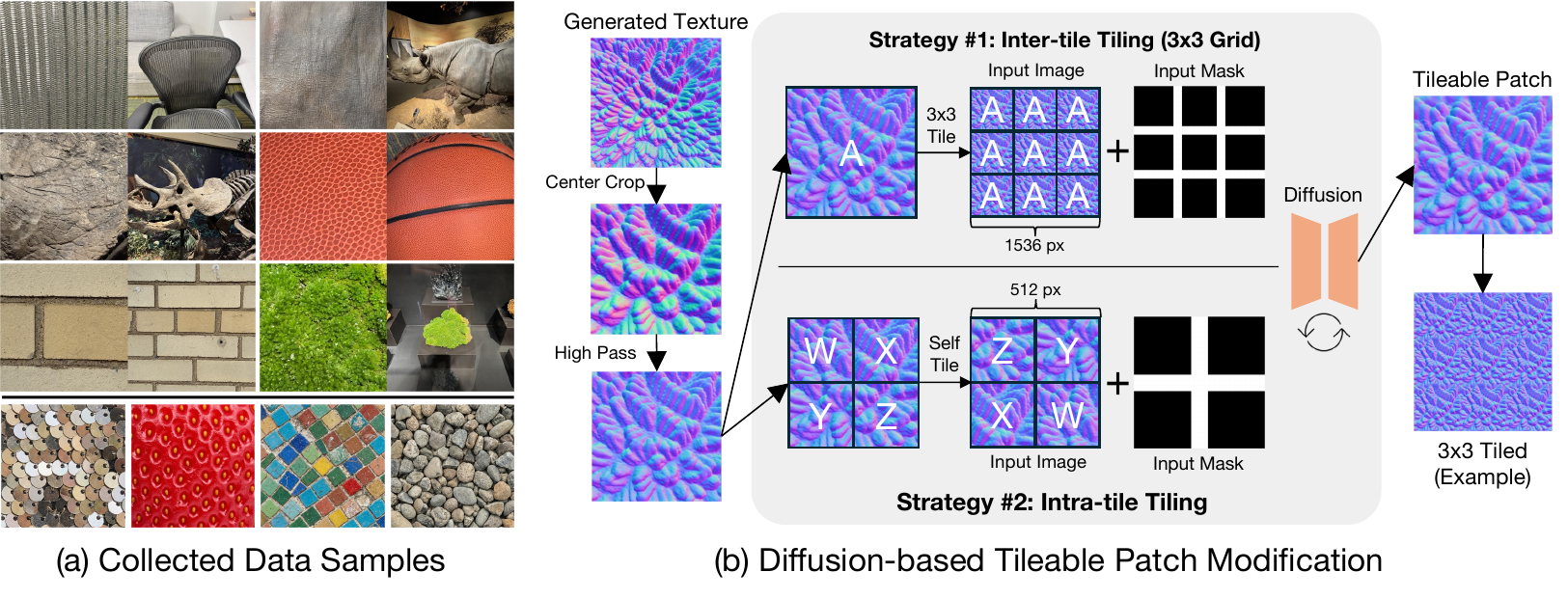}
    \caption{\textbf{Texture data and tileable patch generation.}
    (a) We collect paired real texture images at close-up and zoomed-out scales (top three rows) and augment them with synthetic textures to train our text-to-tactile-texture model (bottom row).
    (b) Generated normal maps are center-cropped and high-pass filtered, then refined through diffusion-based inpainting using either an inter-tile $3\times3$ arrangement or an intra-tile quadrant rearrangement. The resulting patch tiles seamlessly over larger surfaces.}
    \Description{The left panel shows paired close-up and zoomed-out photographs of real textures together with synthetic texture samples. The right panel shows center cropping and high-pass filtering followed by inter-tile or intra-tile diffusion-based refinement to create a seamless tileable normal-map patch.}
    \lblfig{data_tiling}
\end{figure*}

\myparagraph{Texture generation from sensor data.} While the default texture input is specified via a text prompt, our framework also supports tactile sensor data as an alternative texture source. Specifically, we use the TouchTexture dataset introduced in TactileDreamFusion~\cite{gao2024tactile} as a tactile texture library, which contains surface measurements captured by vision-based tactile sensors. We also support processing newly collected tactile data by following the same preprocessing pipeline in \cite{gao2024tactile}, enabling users to incorporate custom tactile measurements captured with tactile sensors (\textit{e.g.}, GelSight~\cite{yuan2017gelsight,wang2021gelsight}). 

\subsection{Tiling}
\lblsec{tiling}
To apply a texture consistently over large surfaces, the generated normal patch must be tileable. Classical texture synthesis methods~\cite{efros2001image,cohen2003wang} rely on minimum-cut or graph-cut algorithms to stitch patches along optimal seam boundaries. While effective for RGB images, these approaches can be computationally expensive as the output size grows, and more fundamentally, cannot modify the patch content. They can only select which portions to retain and where to place seams. For normal maps, this can be problematic, as the minimum cut may discard significant portions of the original texture when ideally only narrow boundary regions should be modified. 

Recent generative approaches~\cite{tiled-diffusion, rodriguez2022seamlessgan, rodriguez2024textile, sartor2024content, fruhstuck2019tilegan, hu2024diffusion} address these limitations by leveraging generative models to synthesize or modify input patches. However, we found that recent training-free diffusion-based methods perform poorly on normal maps due to distribution mismatch. For example, one state-of-the-art approach, Tiled Diffusion~\cite{tiled-diffusion}  often fails to preserve normal map characteristics (see \reffig{tiling_comparison}). Instead, it tends to generate arbitrary objects to achieve tileability, as its primary goal is seamless (and mostly non-repeating) tiling rather than content preservation. It is also optimized for direct text-to-(tileable) image setups and can incur noticeable computational overhead. To this end, we propose a simple combination of inpainting and SDEdit that enables fast synthesis with minimal modification to the original normal map.

\myparagraph{Tileable patch generation.}
Motivated by Tiled Diffusion~\cite{tiled-diffusion} and open-source community efforts~\cite{brick2face,tjm35,levin2025differential}, we combine SDEdit~\cite{sdedit}, which enables minor adjustments while preserving low-level features, with inpainting to enforce continuity in specific boundary regions. We explore two spatial arrangement strategies for this formulation, as illustrated in \reffig{data_tiling}. 
The first strategy, \emph{inter-tile arrangement}, places the input patch in a $3 \times 3$ grid and masks the boundary regions between adjacent tiles. The second, \emph{intra-tile arrangement}, partitions the input patch into a $2 \times 2$ grid of quadrants and swaps them both horizontally and vertically (i.e., point-symmetric rearrangement). Both strategies effectively simulate boundary adjacency by bringing seam regions into close proximity within a single image. Given the rearranged image $I_{\text{swap}}$ and binary inpainting mask $M$ (where $M=1$ indicates regions to inpaint), we apply SDEdit~\cite{sdedit} with Qwen-Image's flow matching scheduler. Specifically, we first add noise to the input by performing $z_{t_0} = (1 - t_0) \cdot \mathcal{E}(I_{\text{swap}}) + t_0 \cdot \epsilon, $ where $\mathcal{E}$ denotes the VAE encoder and $t_0$ corresponds to the denoising strength (or timestep) and $\epsilon \sim \mathcal{N}(0, I)$.

We then iteratively denoise from $t_0$ to $t=0$ using an Euler sampler. At each denoising step $t \rightarrow s$, we blend the denoised latent with the original in the unmasked region by performing $z_s = M \odot z_s^{\text{pred}} + (1 - M) \odot z_s^{\text{orig}}$
where $z_s^{\text{pred}}$ is the model prediction and $z_s^{\text{orig}} = (1 - s) \cdot \mathcal{E}(I_{\text{swap}}) + s \cdot \epsilon$ is the forward-diffused original latent at timestep $s$. We use $t_0 = 0.9$ (denoising strength) with 10 inference steps. We find that the intra-tile arrangement is both faster (since it operates in a smaller single image rather than a $3 \times 3$ grid)
and produces better results. As shown in \reffig{tiling_comparison} and \reftbl{tiling}, compared to Tiled Diffusion, which requires latent wrapping and $\sim$100 sampling steps, our approach generates fewer artifacts while being substantially faster.

In practice, many generated normal-map patches contain undesired low-frequency geometric variations (\textit{e.g.}, global tilts or smooth warping) as well as outlier normals, which complicate seamless tiling. To mitigate this issue, we first center-crop each generated patch from $1024^2$ to $512^2$ resolution and apply a per-channel high-pass filter to remove low-frequency components. Specifically, given a normal map $N = (N_x, N_y, N_z)\in[-1,1]^{H\times W\times 3}$, we filter each channel independently in the Fourier domain:
\begin{equation}
\mathrm{HP}(X) \;=\; \mathcal{F}^{-1}\!\big(\mathcal{F}(X)\odot G_{\mathrm{hp}}\big), 
\quad
G_{\mathrm{hp}}(u,v)=1-\exp\!\left(-\frac{d(u,v)^2}{2\sigma_f^2}\right),
\end{equation}
where $\mathcal{F}$ denotes the 2D FFT, $d(u,v)$ is the distance to the frequency origin, and $\sigma_f$ controls the cutoff frequency. Since high-pass filtering removes the DC component, we restore the flat-normal baseline by adding a constant offset to the $z$-channel and renormalize the normals per pixel:
\begin{equation}
N^{\mathrm{hp}}(p)=\frac{(\mathrm{HP}(N_x),\,\mathrm{HP}(N_y),\,\mathrm{HP}(N_z)+1)}{\left\|(\mathrm{HP}(N_x),\,\mathrm{HP}(N_y),\,\mathrm{HP}(N_z)+1)\right\|_2}.
\end{equation}
This preprocessing yields a cleaner, more uniform patch that reduces global bias and distortion, making subsequent tiling and seamless boundary synthesis easier. We report an ablation in \reffig{tiling_comparison}.

\myparagraph{Texture displacement map.}
To integrate the tiled texture into the base geometry, we integrate the tileable normal texture map into a displacement map.
Let $D_{\text{disp}} \in \mathbb{R}^{H_t \times W_t}$ denote the resulting displacement field.
For each vertex $\mathbf{v}$ on the mesh within the designated textured region, we displace it along its surface normal $\mathbf{n}$ as
$\mathbf{v}' = \mathbf{v} + \lambda \, D_{\text{disp}}(\mathbf{u}) \, \mathbf{n}$,
where $\mathbf{u}$ denotes the vertex’s UV texture coordinate and $\lambda$ controls the displacement magnitude. The designated region is restricted to a user-defined segmentation mask on the mesh, which is by default inferred from the object part specified in the text prompt through the offline segmentation model SAM3~\cite{sam3}. To support finer control, we allow users to interactively refine the segmentation mask by adding or subtracting regions through direct clicking.
This design enables independent control of texture type, scale, and spatial extent, without altering the underlying global geometry.

\subsection{Fabrication-Aware Base Flattening and Braille Placement}
\lblsec{braille}
To provide semantic grounding for the generated geometry, we integrate braille labels onto the base plate.
Because braille readability is highly sensitive to submillimeter geometric variation, the support surface must be strictly planar and horizontal.
However, monocular depth estimation does not recover geometry at this scale, and lifting geometry from predicted depth may introduce unevenness or global tilt in the base plate.
We therefore treat base flattening as a prerequisite for braille generation and enforce a fabrication-aware planar support surface.

Specifically, given the textured mesh, we identify the set of vertices belonging to the base plate using foreground/background segmentation and compute their average height.
All base plate vertices are then projected onto the corresponding horizontal plane, producing a base surface with zero height variation and no global tilt.
In addition, we remove intermediate boundary offsets introduced during geometry lifting, ensuring that both object relief and braille annotations rest directly on a single outer base. Let $\mathcal{V}_{\mathrm{base}}$ denote the set of base plate vertices.
We compute the average base height
$h_{\mathrm{base}} = \frac{1}{|\mathcal{V}_{\mathrm{base}}|} \sum_{\mathbf{v} \in \mathcal{V}_{\mathrm{base}}} v_z$.
Each base vertex $\mathbf{v} = (x, y, z)$ is then projected as $\mathbf{v}' = (x, y, h_{\mathrm{base}})$.
This operation modifies only $z$-coordinates and preserves mesh connectivity by construction.
Additional details on topology preservation, mesh closure for printing, the texture displacement parameter $\lambda$, and texture transitions at boundaries are in \refapp{fabrication_details_supp}.

With a flat and stable base plate, we integrate braille annotations into the tactile graphic.
Input text is converted into braille dots using a standard lookup table and placed on designated flat regions of the base plate.
We follow established U.S.\ and international standards for braille dot size, height, and spacing to ensure tactile readability.
Braille placement is explicitly constrained to regions free of geometric relief and texture displacement, preventing interference during tactile exploration.
The enforced flat base guarantees consistent braille height, uniform spacing, and stable tabletop placement, enabling comfortable two-handed exploration.

\subsection{Interactive Control and Fabrication}
\lblsec{interactive-control}
Our system supports multiple levels of user control.
In the simplest mode, users provide a single text prompt and obtain a fully generated tactile graphic.
Advanced users may optionally specify texture prompts, select object regions for texture application, and adjust texture scale.
These interactions operate exclusively at the texture synthesis stage, leaving the global geometry and braille layout unchanged.
The appendix includes a screen recording of our online demo illustrating these interactive controls. 

The final output is a watertight, manifold mesh with a flat base and enforced minimum feature thickness, suitable for direct 3D printing.
All physical outputs in this work are fabricated using \emph{high-resolution SLA resin 3D printing}, which provides the sub-millimeter geometric accuracy required to faithfully reproduce fine-grained tactile textures and standard-compliant braille features. 
Specifically, we use a Formlabs Form~4 SLA printer for all physical examples.
This fabrication choice enables rapid production of tactile graphics that are portable, durable, and optimized for reliable touch-based exploration.

\section{Experiments}
We conduct evaluations to validate each module and the overall pipeline. Since a core motivation of this work is to aid blind and low-vision (BLV) individuals, we include a dedicated user study with target users in~\refsec{user-study}.

\myparagraph{Implementation Details.} 
Our modular pipeline combines several off-the-shelf components with a fine-tuned texture model. 
For reproducibility, we mainly use the open-source Qwen-Image series~\cite{qwen-image}. 
Base geometry is generated using Qwen-Image-Edit conditioned on both the text prompt and a marble plate template. 
The texture module is obtained by fine-tuning a text-to-image variant of Qwen-Image on a mixture of captured and synthetic texture images using the AdamW optimizer~\cite{adamw} (\refsec{texture-gen}). 
For tiling we use the base diffusion model without additional training.
We compare against Nano Banana Pro~\cite{nanobanana-pro} as a proprietary baseline. 
Both Qwen-Image-Edit and the texture model optionally use a 4-step distilled variant via LoRA trained with Distribution Matching Distillation~\cite{dmd}. 
Monocular geometry and normals are estimated using MoGE v2~\cite{moge-2}, and object regions are obtained with SAM3~\cite{sam3}. 
Additional implementation and training details are provided in \refapp{impl_detail_supp}.

\subsection{Quantitative Evaluation}
Evaluating text-to-tactile graphics is challenging because no standard metrics capture both geometric relief and tactile surface texture, and no prior method covers the full geometry, texture, and braille pipeline. 
We therefore evaluate each module independently, and compare end-to-end outputs with recent text-to-3D approaches in qualitative evaluation (\refsec{qualitative-eval}) and the user study (\refsec{user-study}).

\myparagraph{Base Geometry Generation.}
We evaluate the base geometry generation module on 50 diverse prompts, comparing against Qwen-Image-Edit~\cite{qwen-image} (4-step and 40-step) and Nano Banana Pro, with and without plate template conditioning. 
We detect plate regions using SAM3 and estimate plate heights with MoGE v2. 
A stable base plate is critical for fabrication: large height variation leads to inconsistent plate thickness across samples and uneven surfaces that may occlude relief details. 
To quantify plate stability, for each image $i$ we compute the mean and standard deviation of plate heights within the detected plate region, denoted by $\mu_i$ and $\sigma_i$.
We report the average plate height $\bar{\mu}$ and total plate-height variation on the standard-deviation scale:
$\sigma_{\text{total}} = \sqrt{\frac{1}{N}\sum_i \sigma_i^2 + \mathrm{Var}(\mu_i)}$.
We also report CLIP score~\cite{clip} to measure the semantic alignment between the generated base geometry and the input text prompt. As shown in Table~\ref{tbl:t2i}, plate conditioning consistently improves CLIP scores while stabilizing plate geometry: $\bar{\mu}$ remains tightly around $0.34$--$0.36$ and $\sigma_{\text{total}}$ is significantly reduced across models. 
Without plate conditioning, Nano Banana Pro fails to detect the plate in 6/50 samples, whereas all plate-conditioned variants achieve 100\% detection.

{%
\renewcommand{\thefootnote}{\ifcase\value{footnote}\or
*\or$\dagger$\or$\ddagger$\else\arabic{footnote}\fi}
\begin{table}[t]
    \centering
    \caption{\textbf{Quantitative evaluation.}
    From left to right: \textbf{geometry analysis}\textsuperscript{*} shows plate conditioning improves semantic alignment and plate stability;
    \textbf{texture quality}\textsuperscript{$\dagger$} shows fine-tuning improves consistency and high-frequency detail; and
    \textbf{tileability}\textsuperscript{$\ddagger$} shows our intra-tile method achieves the best seam quality and runtime.}
    \lbltbl{quantitative}
    \lbltbl{t2i}
    \lbltbl{t2texture}
    \lbltbl{tiling}
    \lbltbl{tiling_merged}
    \scriptsize
    \begin{minipage}[t]{0.31\textwidth}
        \centering
        \setlength{\tabcolsep}{2pt}
        \resizebox{\linewidth}{!}{%
            \begin{tabular}{@{}lcccc@{}}
                \toprule
                \multirow{2}{*}{Method} & \multirow{2}{*}{\makecell{Plate\\Cond.}} & \multirow{2}{*}{CLIP$\uparrow$} & \multicolumn{2}{c}{Plate Height} \\
                \cmidrule(l){4-5}
                 & & & Avg. & SD $\downarrow$ \\
                \midrule
                QIE-4  & \xmark & 34.82 & 0.181 & 0.131 \\
                QIE-4  & \cmark & 36.64 & 0.337 & \textbf{0.122} \\
                QIE-40 & \xmark & 36.11 & 0.305 & 0.205 \\
                QIE-40 & \cmark & \underline{36.91} & 0.350 & \underline{0.123} \\
                \midrule
                NBP & \xmark & 36.83 & 0.467 & 0.188 \\
                NBP & \cmark & \textbf{37.66} & 0.365 & 0.137 \\
                \bottomrule
            \end{tabular}%
        }
    \end{minipage}\hfill%
    \begin{minipage}[t]{0.34\textwidth}
        \centering
        \setlength{\tabcolsep}{2pt}
        \resizebox{\linewidth}{!}{%
            \begin{tabular}{@{}lccc@{}}
                \toprule
                Method & CLIP$\uparrow$ & \makecell{Patch\\Self-Sim.$\downarrow$} & \makecell{HF\\Ratio$\uparrow$} \\
                \midrule
                QI-4    & 27.14 & 0.496 & 0.305 \\
                QI-40   & 27.68 & 0.461 & 0.314 \\
                Ours-4  & \textbf{28.98} & 0.412 & 0.396 \\
                Ours-40 & 28.43 & \underline{0.400} & \textbf{0.408} \\
                \midrule
                NBP     & \underline{28.62} & \textbf{0.395} & \underline{0.402} \\
                \bottomrule
            \end{tabular}%
        }
    \end{minipage}\hfill%
    \begin{minipage}[t]{0.33\textwidth}
        \centering
        \setlength{\tabcolsep}{2pt}
        \resizebox{\linewidth}{!}{%
            \begin{tabular}{@{}lccc@{}}
                \toprule
                Method & \makecell{Border\\Conti.$\downarrow$} & \makecell{Grad.\\Conti.$\downarrow$} & \makecell{Time\\(s)} \\
                \midrule
                Resize    & 29.80 & 59.99 & -- \\
                Crop      & 26.10 & 52.82 & -- \\
                + HPF     & 22.84 & 46.53 & -- \\
                ++ TD     & \underline{4.43} & \underline{9.76} & \underline{10.9} \\
                ++ Ours-I & 7.19 & 14.42 & 24.88 \\
                ++ Ours-A & \textbf{1.61} & \textbf{3.19} & \textbf{3.18} \\
                \bottomrule
            \end{tabular}%
        }
    \end{minipage}
\end{table}

\footnotetext[1]{QIE = Qwen-Image-Edit; NBP = Nano Banana Pro.}
\footnotetext[2]{QI-4/40 = Qwen-Image (4/40-step); NBP = Nano Banana Pro. Patch Self-Sim. denotes average patchwise LPIPS, and HF Ratio denotes the proportion of high-frequency normal-map variation.}
\footnotetext[3]{TD = Tiled Diffusion; Ours-I/A = inter-/intra-tile. Border Conti. measures edge discontinuity, and Grad. Conti. measures slope discontinuity.}
}

\myparagraph{Text-to-Texture.}

We evaluate texture generation on 50 diverse prompts, comparing our fine-tuned Qwen-Image model against the base model and Nano Banana Pro. Standard text-to-image models are object-centric: prompting ``avocado skin texture'' often yields an entire round-shaped avocado body rather than a uniform surface pattern. To measure spatial consistency, we generate all textures at 1024$\times$1024 resolution, randomly sample 32 patches of 128$\times$128 pixels, and compute the average pairwise LPIPS (VGG) distance~\cite{lpips}. Lower values indicate more uniform patterns suitable for tiling.
We also compute an FFT-based high-frequency (HF) ratio from estimated normal maps, measuring the fraction of non-DC normal-map variation that remains after applying our Gaussian high-pass normal-map filter, and report CLIP Score to measure alignment with the texture prompt.
As shown in Table~\ref{tbl:t2texture}, our fine-tuned model generally achieves lower LPIPS and higher HF ratio than the base model, indicating improved pattern consistency and a greater concentration of high-frequency normal variation. 
Performance is comparable to Nano Banana Pro while remaining fully reproducible. 
\reffig{baseplate_texture} shows qualitative comparisons and \refapp{eval_details} provides additional evaluation details.

\myparagraph{Tiling.}
We ablate each preprocessing stage on 50 normal maps from our fine-tuned model (4-step), targeting $512\times512$ tileable patches. We compare: (1) plain resize from 1024 to 512, (2) center crop, (3) center crop with high-pass filtering, and (4) diffusion-based tiling.
Seam quality is measured using two metrics (lower is better): border continuity, which measures pixel discontinuity at tile edges, and gradient continuity, which measures slope smoothness across seams. 
As shown in Table~\ref{tbl:tiling}, each preprocessing stage progressively improves seam quality, and our intra-tile tiling approach achieves the best results.
We also compare against the most recent training-free tiling work, Tiled Diffusion~\cite{tiled-diffusion}. 
While it produces visually plausible textures, it often alters the underlying normal map structure, resulting in semantically incorrect tactile patterns. 
Qualitative examples are shown in \reffig{tiling_comparison}. Full metric definitions are provided in \refapp{eval_details}.

\begin{table*}[t]
    \centering
    \caption{\textbf{User preference study.}
    We evaluate the perceptual quality of the generated textures, reporting Likert score ranges 1--5 averaged across each user population group.}
    \lbltbl{user_study}

    \small
    \setlength{\tabcolsep}{10pt}
    \begin{tabular}{@{}lccc@{}}
        \toprule
        \multirow{2}{*}{Population} & \multicolumn{3}{c}{Method} \\
        \cmidrule(lr){2-4}
         & Hunyuan3D & Naive & \textbf{Ours} \\
        \midrule
        Blind and low-vision (BLV) & $3.62 \pm 1.28$ & $3.62 \pm 1.20$ & \textbf{$\bm{3.93 \pm 1.32}$} \\
        Blindfolded sighted (BSP) & $2.97 \pm 1.31$ & $3.02 \pm 1.24$ & \textbf{$\bm{3.35 \pm 1.39}$} \\
        \bottomrule
    \end{tabular}
\end{table*}

\begin{figure}[t]
    \includegraphics[width=\textwidth]{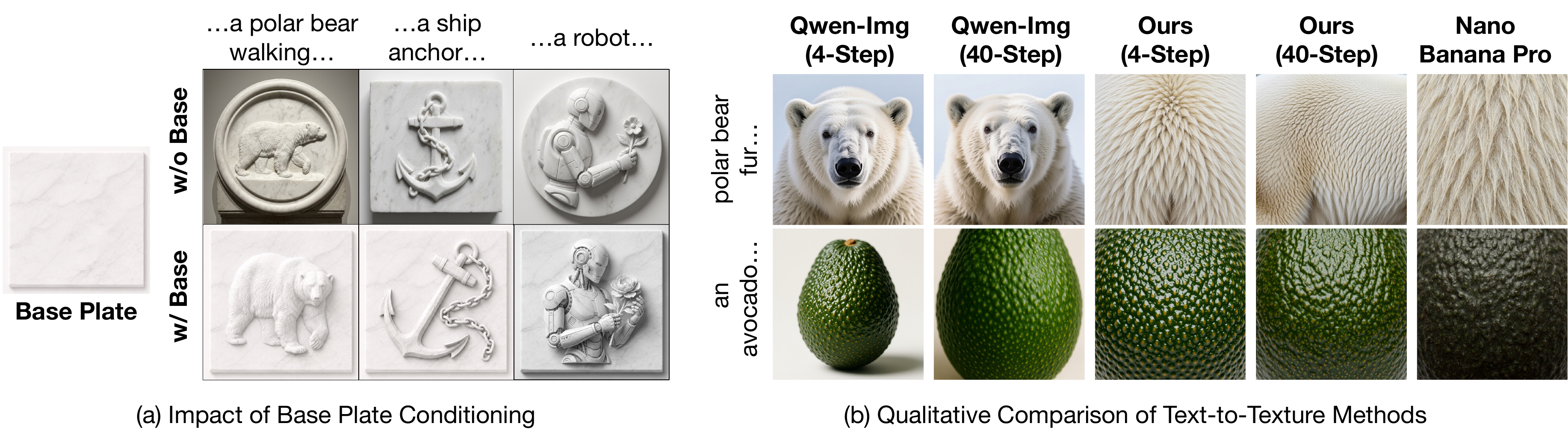}
    \caption{\textbf{Base-plate conditioning and text-to-texture generation.}
    (a) Without base-plate conditioning, generated reliefs exhibit inconsistent, uneven, or missing plates. A fixed base-plate template stabilizes plate geometry and relief height/shape across prompts.
    (b) We compare base Qwen-Image, our fine-tuned variants, and Nano Banana Pro. Our models generate homogeneous, fine-grained texture patterns instead of object-centric images, making them suitable for tiling and tactile synthesis.}
    \Description{The left panel compares relief images generated without and with a fixed base-plate template. The right panel compares polar-bear-fur and avocado textures generated by base Qwen-Image, the fine-tuned models, and Nano Banana Pro.}
    \lblfig{baseplate_texture}

\end{figure}

\subsection{Qualitative Evaluation}
\lblsec{qualitative-eval}
\myparagraph{Baselines for pipeline evaluation.}
Unlike prior approaches that produce tactile graphics on swell paper~\cite{khan2025tactilenet}, our method outputs \emph{fabrication-ready 3D meshes} that jointly encode geometry, surface texture, and braille. 
As this constitutes a new problem setting, we compare our approach against recent state-of-the-art 3D generation methods, including Hunyuan3D~2.1~\cite{hunyuan3d-2-1} and Trellis~2~\cite{trellis2}. 
Since they are designed for object-centric 3D generation, 
they often fail to produce the stable base plate required for tactile graphics (\reffig{baseline_3d_failures}). 
Hunyuan3D may recognize the plate but place it at incorrect depths, while Trellis~2 fails when the plate is present in the input image. 
To enable comparison, we adapt Trellis~2 by removing the plate background and placing the generated relief onto a fixed base plate.
We also include a naive baseline that shares the same base geometry pipeline but omits the explicit text-to-texture and braille modules. Instead, it directly maps the full text prompt to a single relief image, which is then lifted to 3D. 
Qualitative comparisons across all methods are shown in \reffig{baseline_comp}.
Our method achieves the most consistent plate geometry while producing richer and more semantically aligned surface textures.
Additional qualitative results for ablations of the base plate template, comparisons against traditional 2.5D representations, and material- and texture- generation methods are provided in \refapp{supp_additional_comparison_25D}.

\begin{figure*}[t] 
    \centering 
    \includegraphics[width=\linewidth]{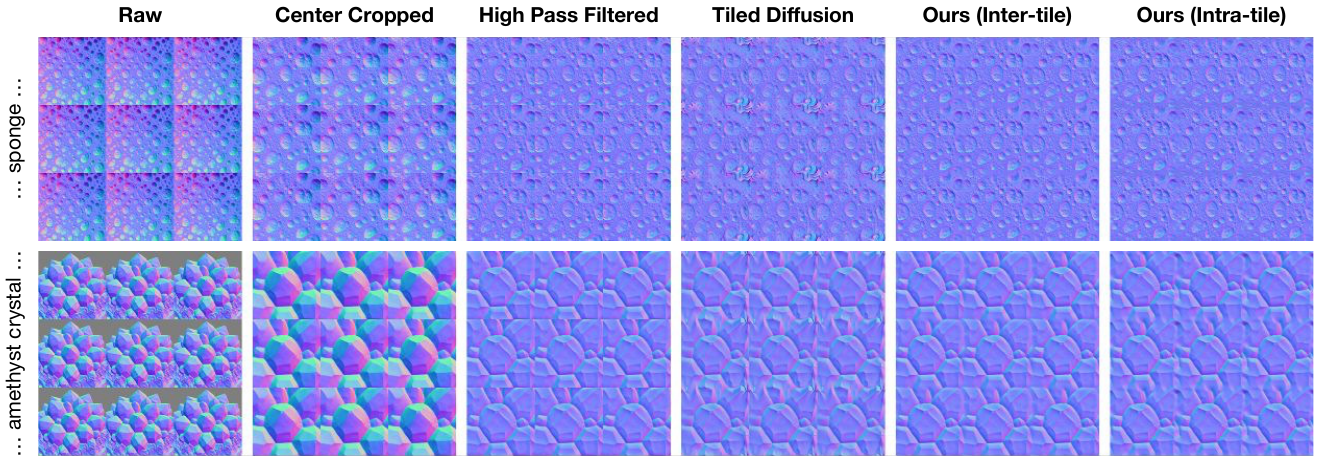} 
    \caption{\textbf{Tiling comparisons.} Our center cropping and high-pass filtering steps make textures more consistent and suitable for application to 3D geometry. Compared to Tiled Diffusion~\cite{tiled-diffusion}, our method produces more seamless textures, with the intra-tile configuration having the best results. Best viewed zoomed in.}
    \Description{Our intra-tile tiling method produces the most seamless and consistent textures.}
    \lblfig{tiling_comparison} 
\end{figure*}

\begin{figure*}[htbp] 
\centering 
\includegraphics[width=\linewidth]{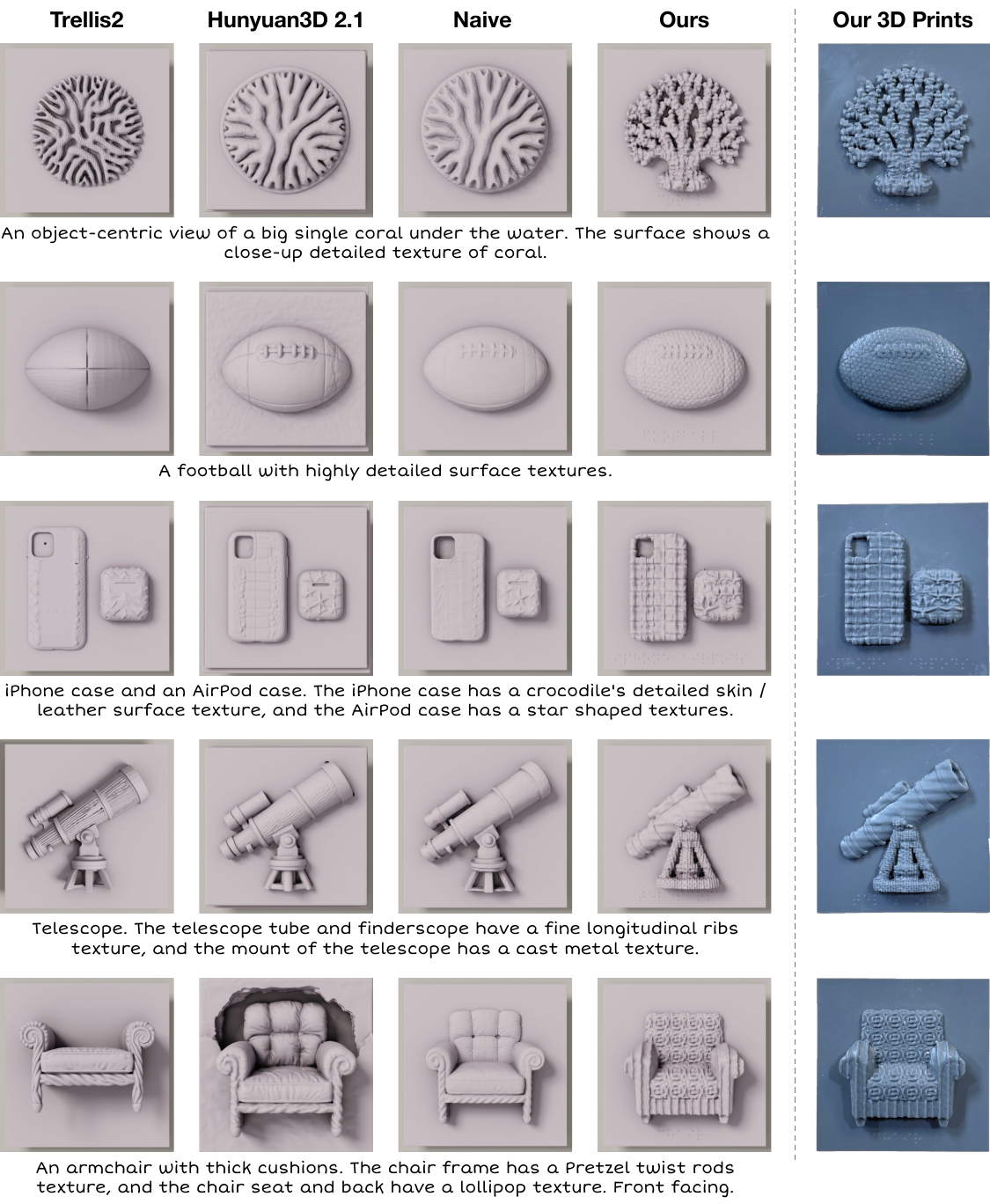} 
\caption{\textbf{Qualitative comparison of 2.5D reliefs.} Given the same text prompt, our method produces consistent plate geometry with readable braille and fine-grained textures aligned with the prompt. 
In contrast, baselines often lack texture detail (overly smooth corals in Row 1), produce non-fabricable surfaces (fragmented surface of telescope in Row 4 Col 1), or generate broken geometry (chair intersecting with the base plate in Row 5 Col 2). Please zoom in for finer details, including the braille.}

\Description{Our method produces physically fabricable tactile graphics with braille annotations alongside consistent, targeted texture placement, such as a phone case with a crocodile-skin texture. In contrast, baseline methods fail to include sufficient texture or result in non-fabricable outputs, such as a chair intersecting the base plate.}
\lblfig{baseline_comp}
\end{figure*}

\subsection{In-Person Perceptual Study}
\lblsec{user-study}
\myparagraph{Study Design.}
We conduct an in-person perceptual study to evaluate the usability and tactile quality of generated 2.5D tactile graphics, in collaboration with a state-run regional library for accessible media.
We fabricate tactile graphics for 16 diverse prompts using three methods: \emph{Hunyuan3D~2.1}, \emph{Ours w/o text-to-texture (Naive)}, and \emph{Ours (full)}.
Outputs from Trellis~2 frequently fail fabrication checks due to non-manifold geometry or thin structures and are therefore excluded from physical evaluation (see~\reffig{baseline_3d_failures}). 
In total, 48 tactile samples are fabricated.

Two groups, \emph{(1) eight blind and low-vision (BLV) participants} with varying vision conditions and braille literacy levels, and \emph{(2) seven blindfolded sighted (BSP) participants} who do not read braille, totaling 15 users, participated.
The study protocol is approved by the appropriate institutional review process.
Each session lasts 45--60 minutes and consists of two tasks. \emph{(1) Object recognition:} BLV participants explore each tactile graphic without prior information and attempt to identify the depicted object, both with and without braille annotations. \emph{(2) Texture quality rating:} All participants are informed of the target texture and rate how well the tactile surface matches their expectation on a 5-point Likert scale.
To mitigate bias, condition labels were not disclosed to participants, the three artifacts per prompt were shown in randomized order, and plate dimensions were standardized so that no condition was identifiable by its form factor.

\myparagraph{Results.}
Our method receives higher ratings than all baselines across both participant groups (\reftbl{user_study}), demonstrating the benefit of explicitly modeling tactile textures.
To verify statistical robustness, we ran one-sided paired $t$-tests over a pool of 240 paired observations (16 prompts $\times$ 15 participants). 
We confirm a statistically significant preference for our method over both Naive and Hunyuan3D baselines (Ours $>$ Naive: mean diff.\,$= 0.32$, $p = 0.008 \ll 0.05$, Cohen's $d = 0.16$; Ours $>$ Hunyuan3D: mean diff.\,$= 0.34$, $p = 0.006 \ll 0.05$, Cohen's $d = 0.16$).
Adding braille increases the object recognition accuracy from $\sim20\%$ to nearly perfect, confirming that the printed braille is readable and effective. 
We additionally compare our 3D-printed outputs against expert-designed swell-paper graphics in \refapp{artist_swell_comparison}.
Please refer to \refapp{extended_user_study} for more in-depth qualitative analysis.

\section{Discussion}
We have presented an automated method for generating 3D-printable tactile graphics directly from text prompts. 
Unlike traditional embossed or swell paper graphics, our approach produces expressive 2.5D representations that integrate geometry, tactile textures, and braille.
User studies and quantitative analysis demonstrate the effectiveness of our generative graphics for physical fabrication and tactile exploration. 
We hope our work enables more accessible and personalized tactile graphics for the blind and low-vision community and inspires new research directions in accessibility for generative models. 

\myparagraph{Limitations.}
Object recognition from touch alone is inherently difficult. Our $\sim$20\% recognition rate without braille is consistent with prior evidence that passive touch poorly recovers global shape~\cite{miller2022evidence,kuroki2025passive}. 
With standard-compliant braille, however, our generated artifacts function as a self-sufficient platform for exploring fine surface details, experiencing otherwise inaccessible materials, and enabling customized design and artistic expression through touch.
Additionally, our in-person perceptual study involves 15 participants (8 BLV, 7 BSP) recruited from the local community; while the results are statistically significant, expanding to a larger and more geographically diverse BLV population remains important future work.

\myparagraph{Acknowledgment. }
We thank Amin Mirzaee, Maxwell Jones, and Nupur Kumari for their helpful comments and discussion. We are also grateful to Sheng-Yu Wang for proofreading the draft. 
We appreciate the generous help from the Library of Accessible Media for Pennsylvanians (LAMP) and VisAbility Pittsburgh for recruiting blind and low-vision users for our studies. 
The project is partially supported by an NVIDIA Academic Grant, Amazon Faculty Research Award, Cisco Research, Sloan Foundation, the Packard Foundation, and the IITP grant funded by the Korean Government (MSIT) (No. RS-2024-00457882, National AI Research Lab Project).  
Ruihan Gao is supported by the A*STAR National Science Scholarship (Ph.D.).
Joonghyuk Shin and Jaesik Park are supported by the NRF grant (No. RS-2024-00405857, 50\%) and IITP grant (No. RS-2021-II211343, 5\% and No. RS-2024-00509257, 45\%) funded by the Korea government (MSIT).

\bibliographystyle{splncs04}
\bibliography{main}
\clearpage
\setcounter{table}{0}
\setcounter{page}{1}
\setcounter{figure}{0}
\setcounter{equation}{0}
\renewcommand{\thefigure}{A\arabic{figure}}
\renewcommand{\thetable}{A\arabic{table}}
\appendix
\section*{Appendix Overview}
This appendix provides additional technical, implementation, and evaluation details that support the main paper. We first describe implementation specifics, including prompt templates, dataset construction, braille generation, interface of the interactive demo, and mesh processing and fabrication details in \refsec{impl_detail_supp}. We then present details of evaluation metrics and analyze failure cases of object-centric 3D baselines in \refsec{supp_metrics}. Next, we provide extended qualitative analysis from the in-person user study, highlighting perceptual and experiential insights not captured by quantitative metrics in \refsec{extended_user_study}. Finally, we report additional evaluation results, qualitative comparison to traditional 2.5D representations in \refsec{supp_additional_comparison_25D}, a perceptual comparison against expert-designed swell-paper graphics sourced from the APH Tactile Graphic Image Library in \refsec{artist_swell_comparison}, and additional qualitative comparisons against material- and texture- generation baselines in \refsec{material_gen_comparison}.

\section{Implementation Details}
\lblsec{impl_detail_supp}
\subsection{Prompt Templates}
With advanced text-to-image (T2I) models capable of capturing subtle nuances in input prompts, effective prompt engineering has become increasingly important. Through empirical trials, we identified a set of robust prompt templates that consistently yield high-quality results across diverse scenarios. Below, we detail the fixed prompt templates used for base geometry, texture synthesis, and tiling.

\myparagraph{Base Geometry Generation.}
For the base relief generation, we wrap the user's input object description with a fixed style prefix and suffix to enforce the marble relief aesthetic and correct camera viewpoint. The full prompt is constructed as: \texttt{\small STYLE\_PROMPT\_PREFIX} + \textit{[Input Object]} + \texttt{\small STYLE\_PROMPT\_SUFFIX}.

\begin{itemize}
    \item \textbf{Prefix:} ``\texttt{\small A white marble stone relief sculpture of:}''
    \item \textbf{Suffix:} ``\texttt{\small Depicted as a detailed relief sculpture carved from white marble, with well-defined textures and crisp, well-defined edges, in a monochrome white-to-gray color scheme. The sculpture is carved and formed directly on top of the provided marble plate, using the plate as the base material. The relief is gently but clearly convex, rising higher above the plate surface to create readable volume and form, while remaining a restrained relief rather than a fully detached sculpture. It is never engraved inward, recessed, or cut into the stone. Simple, centered composition with top-down lighting that emphasizes surface depth and height. Viewed from directly above, fully confined within the marble plate.}''
\end{itemize}

\myparagraph{Texture Generation.}
For the text-to-tactile-texture module, we use the following template to enforce the specific camera angle and lighting conditions required for tileable textures. The \textit{\{texture description\}} placeholder is replaced by the user's input or the VLM-generated description.

\begin{itemize}
    \item \textbf{Template:} ``\texttt{\small A texture of \textit{\{texture description\}}, extreme close-up, plain lighting, straight-on view}''
\end{itemize}

\myparagraph{Tiling Generation.}
For the diffusion-based tiling refinement step, we utilize specific positive and negative prompts to encourage seamlessness in the normal map domain while suppressing grid artifacts.

\begin{itemize}
    \item \textbf{Positive Prompt:} ``\texttt{\small a highly detailed surface normal map (normal map color scheme) smooth, continuous, repetitive patterns}''
    \item \textbf{Negative Prompt:} ``\texttt{\small seams, visible seams, tiled, nxn tiles}''
\end{itemize}

\subsection{Dataset Preparation and Training details for the Text-to-Tactile-Texture Module}

In this section, we describe the data generation and training process for the text-to-tactile-texture module described in \refsec{texture-gen}.

To better align the model with our target distribution, we curated a small but high-quality dataset by manually capturing approximately 70 real-world texture images. These images were provided to a VLM (\textit{i.e.,} the Gemini 3 API) as in-context examples, which were then used to generate approximately 2,500 additional prompts adhering to our desired stylistic constraints. Empirically, we found that conditioning the VLM on our images allowed it to better capture the underlying distribution of our dataset, producing prompts that more closely matched our intended texture styles compared to na\"ive prompt generation. To further enforce the desired distribution, we explicitly instructed the model to include keywords such as ``close-up'', ``plain lighting'', and ``straight-on view''.

We then fine-tune Qwen-Image with LoRA (rank 32) for 3,000 steps using AdamW optimizer~\cite{adamw}, with a learning rate of $10^{-4}$ and a batch size of 32 on 8 NVIDIA A100 GPUs. The training process takes approximately 7 hours. After training, as mentioned in the main text, we accelerated the sampling process by adopting a plug-and-play 4-step distilled LoRA. During inference, our model generates a high-quality RGB texture image, which is then fed into an off-the-shelf monocular normal estimator (MoGE v2~\cite{moge-2}) and subsequently post-processed as described in the main text.
\lblsec{supple_dataset_training_details}

\subsection{Braille Generation}
Given a textual annotation to be rendered in braille (by default, the object name), we first convert the input text into standard braille cells and dot patterns. Following established braille formatting standards~\cite{authority2011guidelines}, we adopt a dot height of 0.5\,mm, dot diameter of 1.5\,mm, intra-cell spacing of 2.5\,mm, and inter-cell spacing of 6\,mm. Each braille dot is modeled as a smooth dome-shaped geometry to facilitate comfortable tactile reading.

We compute the 2D locations of all braille dots on the base plate, project the resulting braille mask onto the mesh, and displace the corresponding vertices according to the predefined dome profile. 
To improve fabrication robustness, each dot is slightly enlarged to provide additional tolerance for SLA 3D printing. 
Owing to the enforced flat base plate, sufficient planar area is reserved for braille placement. In all examples, a single braille annotation is rendered per object and positioned within the bottom 15\% of the plate, with center alignment for consistent layout.

\subsection{Interactive demo}
\reffig{demo_screenshot} presents the interactive demo interface used to generate tactile graphics with the proposed system. The demo exposes the full generation pipeline, allowing users to specify object descriptions, assign tactile textures to individual regions via either text- or click-based segmentation, and generate tileable textures from text prompts or sensor-captured data. The resulting textured 2.5D relief can be previewed and exported as a fabrication-ready mesh, illustrating how the system supports interactive, part-aware, and multimodal tactile graphic design.
\begin{figure*}[tb] 
\centering 
\begin{subfigure}[T]{0.49\linewidth}
    \centering
    \includegraphics[width=\linewidth]{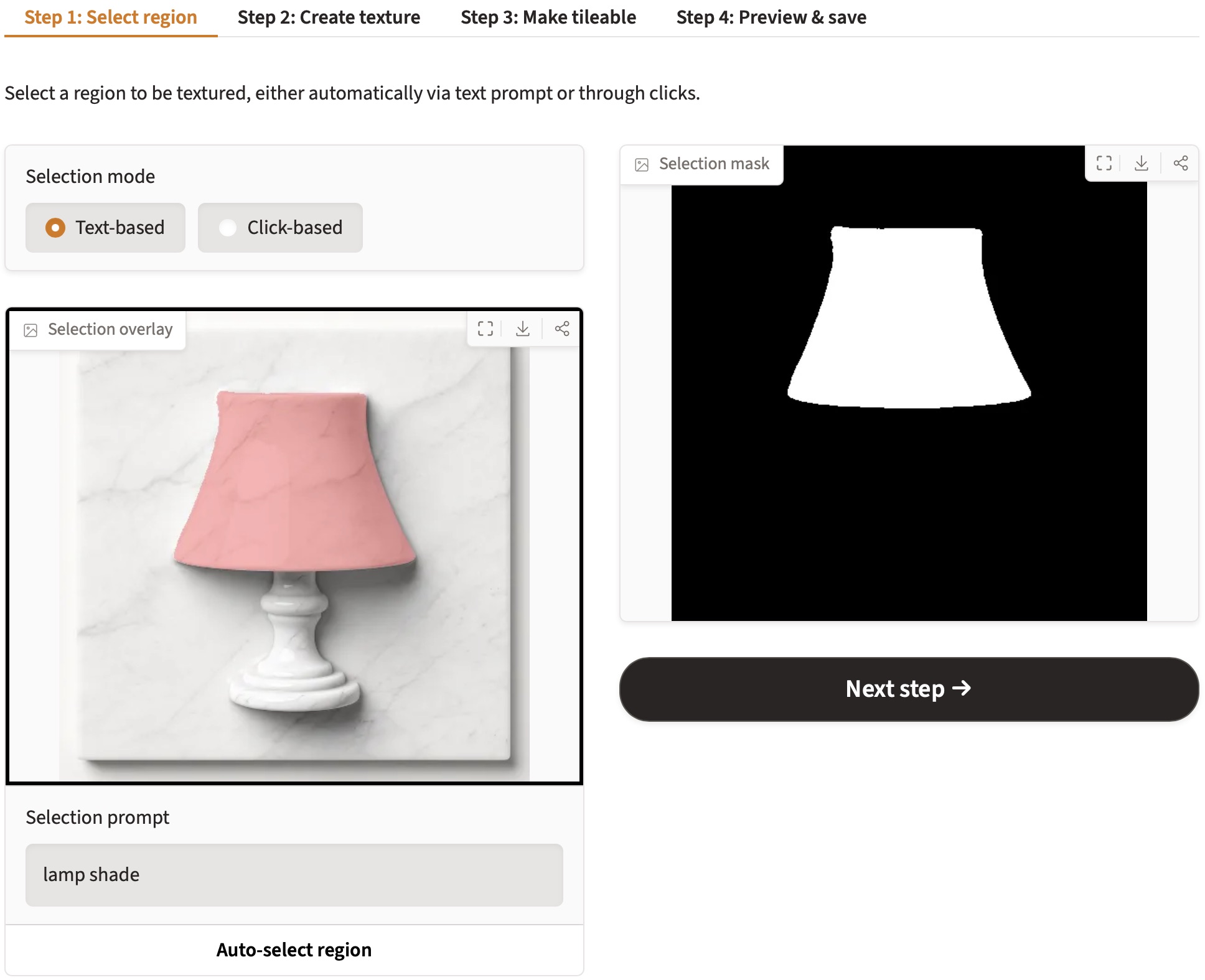}
\end{subfigure}%
\hfill%
\begin{subfigure}[T]{0.49\linewidth}
    \centering
    \includegraphics[width=\linewidth]{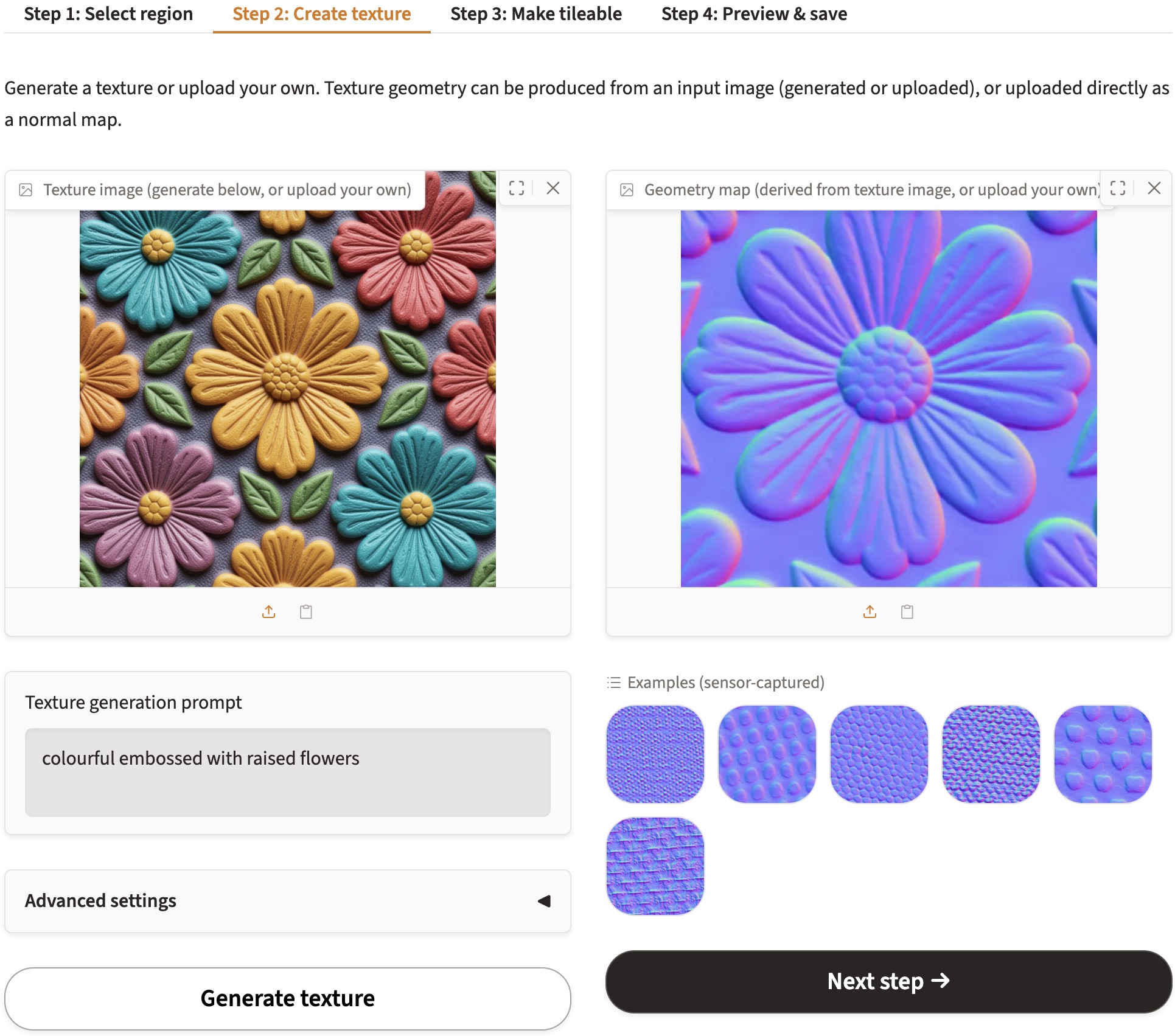}
\end{subfigure}

\begin{subfigure}[T]{0.49\linewidth}
    \centering
    \includegraphics[width=\linewidth]{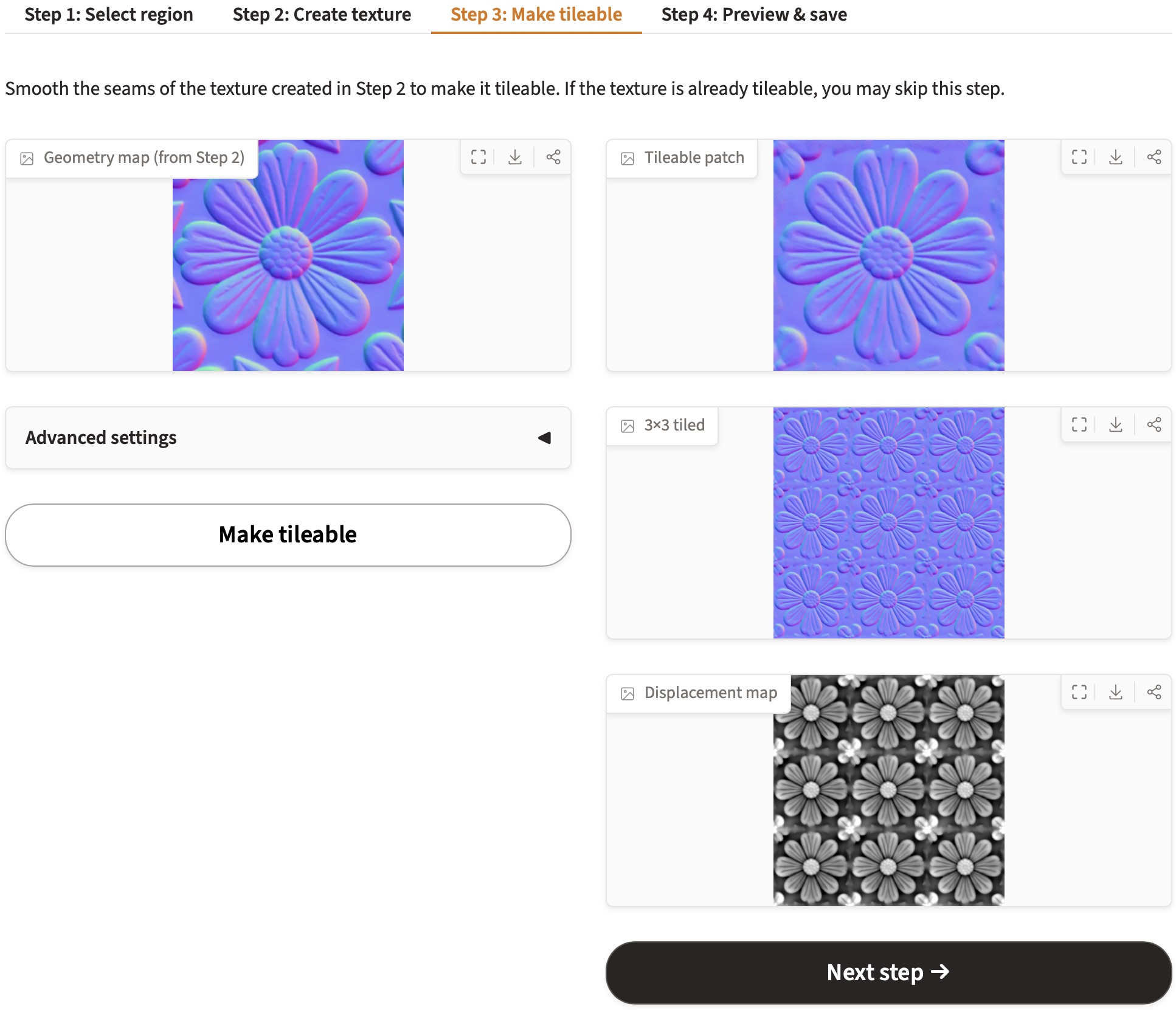}
\end{subfigure}%
\hfill%
\begin{subfigure}[T]{0.49\linewidth}
    \centering
    \includegraphics[width=\linewidth]{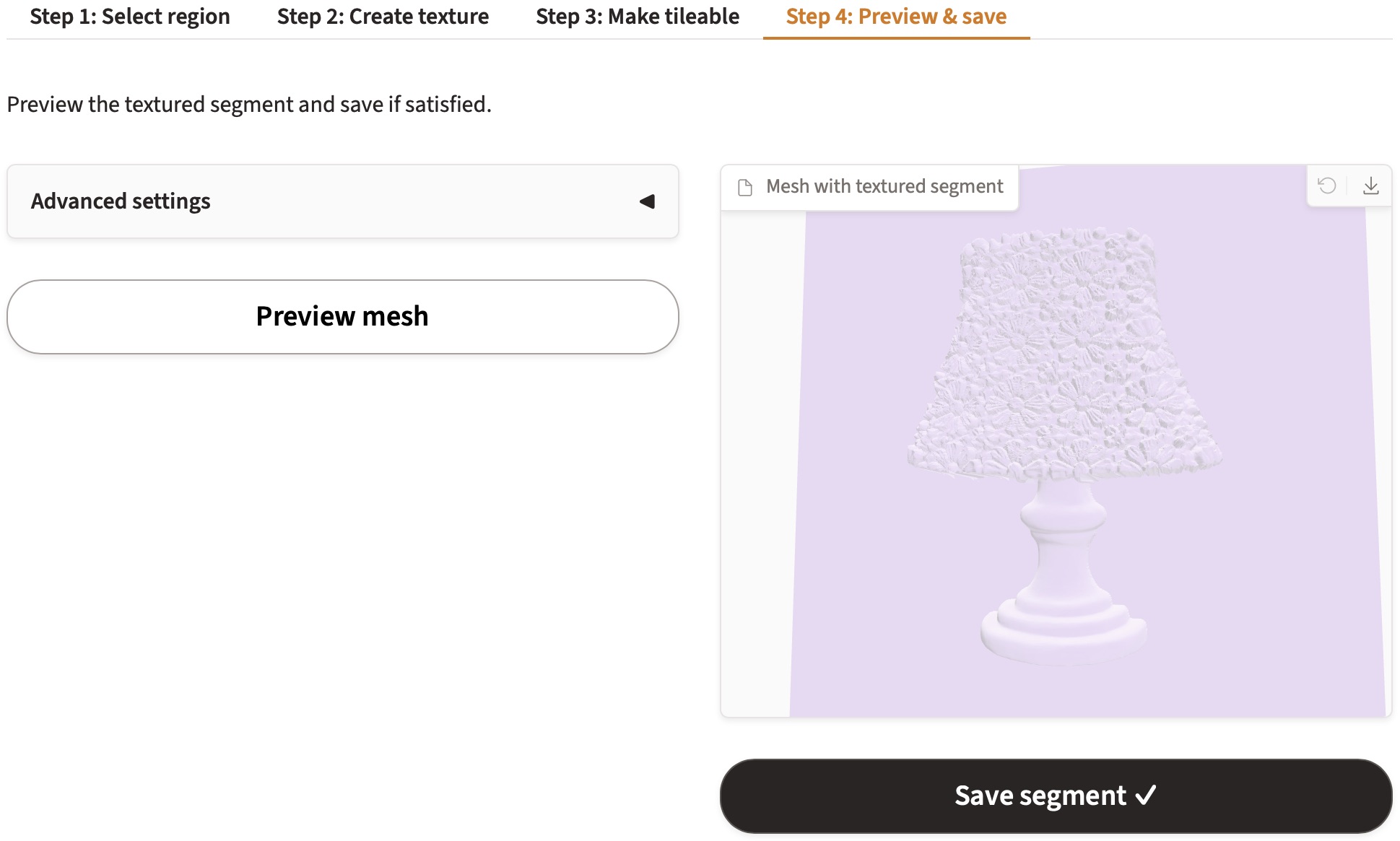}
\end{subfigure}
\caption{\textbf{Interactive demo interface for generative tactile graphics.} 
The web-based demo supports end-to-end tactile graphic generation from text prompts, including region selection, texture generation, tiling, and preview of the resulting 2.5D relief. Users can assign different tactile textures to object parts via text input or interactive segmentation, preview the textured relief geometry, and export the final mesh for 3D printing, via a step-by-step process.}
\lblfig{demo_screenshot} 
\end{figure*}

\subsection{Mesh Processing and Fabrication Details}
\lblsec{fabrication_details_supp}
We provide additional details on the mesh processing and fabrication pipeline described in \refsec{braille} and \refsec{interactive-control} of the main paper.

\myparagraph{Topology Preservation during Base Flattening.}
As described in the main paper, the base flattening step projects all base plate vertices $\mathcal{V}_{\mathrm{base}}$ onto the horizontal plane at height $h_{\mathrm{base}}$. Crucially, this operation only translates vertex $z$-coordinates; no edges, faces, or vertex connectivity are modified. Because the underlying representation is a 2.5D height field with fixed $(x,y)$ vertex positions, the flattening preserves mesh connectivity and avoids fold-over or self-intersection by construction. Each base vertex is moved independently along $z$, so no triangle is cut, rewired, or degenerated during this step.

\myparagraph{Mesh Closure and Fabrication Validation.}
After base flattening and texture displacement, the resulting mesh is an open surface (a height field). To produce a watertight solid suitable for 3D printing, we add vertical side walls along the mesh boundary and a planar bottom slab, closing the exterior into a manifold volume. Standard mesh cleanup operations (duplicate vertex removal, degenerate face elimination, etc.) are then applied. All 48 fabricated samples (16 prompts $\times$ 3 methods) pass the Formlabs slicer's manifold and watertight checks, confirming that the pipeline reliably produces print-ready geometry.

\myparagraph{Texture Displacement Parameter $\lambda$.}
The displacement magnitude $\lambda$ in the texture integration equation $\mathbf{v}' = \mathbf{v} + \lambda \, D_{\text{disp}}(\mathbf{u}) \, \mathbf{n}$ (main paper, \refsec{tiling}) controls the physical relief depth of applied textures. In practice, $\lambda$ is specified as an absolute height and is tuned within the range of approximately $3$\,mm (for fine, subtle textures) to $6$\,mm (for coarse, pronounced textures), depending on the desired tactile intensity. This range was determined empirically based on (1)~the minimum feature size reliably resolved by the SLA printer ($\sim$50\,$\mu$m), and (2)~pilot feedback from BLV participants, who indicated a preference for balanced relief amplitude rather than maximal geometric exaggeration (\refsec{extended_user_study}). The parameter is user-adjustable in the interactive demo (\refsec{impl_detail_supp}).

\myparagraph{Texture Transitions at Part Boundaries.}
Because textures are applied independently per segmentation mask, small height discontinuities can occur at the boundaries between differently textured parts (\textit{e.g.,} the seat and frame of a chair, or the cap and stalk of a mushroom). These transitions are expected by design and serve as tactile segmentation cues that help users distinguish object regions through touch, as confirmed by participants in our user study (\refsec{extended_user_study}). Within each textured region, the displacement field is tileable and continuous; the only discontinuities occur at the explicitly defined part boundaries. Across our fabricated samples, we did not observe sharp cracks or detached seams at either base plate or part boundaries, and no participant reported that such transitions impaired tactile perception.

\section{Additional Evaluation Results and Details}
\lblsec{supp_metrics}
\begin{table}[t]
\centering
\setlength{\tabcolsep}{5pt}
\caption{\textbf{Additional metrics for base geometry generation baselines.} Qwen refers to the same Qwen-Image-Edit model used throughout.}
\label{tbl:base_geometry_metrics_supp}
\begin{tabular}{lccccc}
\toprule
\multirow{2}{*}{Baseline} & \multirow{2}{*}{Plate cond.} & \multirow{2}{*}{FG $\downarrow$} & \multicolumn{3}{c}{Plate Height Std. Dev.$\downarrow$} \\
\cmidrule(lr){4-6}
 &  &  & Within & Between & Total \\
\midrule
Qwen (4-Step)  & \xmark & 0.343 & 0.097 & 0.087 & 0.131 \\
Qwen (4-Step)  & \cmark & 0.267 & 0.052 & 0.110 & 0.122 \\
Qwen (40-Step) & \xmark & 0.380 & 0.090 & 0.184 & 0.205 \\
Qwen (40-Step) & \cmark & 0.244 & 0.060 & 0.107 & 0.123 \\
\midrule
Nano Banana Pro      & \xmark & 0.323 & 0.065 & 0.176 & 0.188 \\
Nano Banana Pro      & \cmark & 0.268 & 0.071 & 0.117 & 0.137 \\
\bottomrule
\end{tabular}
\end{table}

\myparagraph{Base Geometry Generation.}
As shown in \reftbl{base_geometry_metrics_supp}, to assess the stability of the generated base plate, we decompose the total plate-height variation into within-sample and between-sample components, and report all values on the standard-deviation scale. For each image $i$, let $\mu_i$ and $\sigma_i$ denote the mean and standard deviation of plate heights within the detected plate region. The within-sample variation, $\sqrt{\frac{1}{N}\sum_i \sigma_i^2}$, measures how flat and uniform each generated plate is, where a lower value is essential for stable braille placement and preserving low-relief features. The between-sample variation, $\sqrt{\mathrm{Var}(\mu_i)}$, measures how consistent the average plate height is across different prompts and seeds. The total variation is computed as $\sigma_{\text{total}} = \sqrt{\frac{1}{N}\sum_i \sigma_i^2 + \mathrm{Var}(\mu_i)}$. We also report the average foreground area, calculated from the SAM3 mask. Due to the diversity of prompts, the foreground area is not inherently better when larger or smaller; nevertheless, we generally observe slightly smaller foreground regions when a plate is used, since generated content is constrained to remain within the plate boundary.

\lblsec{eval_details}

\myparagraph{Text-to-Texture.}
To assess the frequency distribution of generated tactile textures, we report the high-frequency ratio (HF ratio), which measures the fraction of non-DC normal-map variation retained after applying our Gaussian high-pass normal-map filter.
Given a normal map, we first compute its total spectral energy, $E_{\text{total}}$, in the frequency domain using Parseval's theorem, explicitly excluding the DC component ($k=0$) to focus on surface variations. We then apply a Gaussian high-pass filter with a spatial wavelength cutoff of $\lambda = 120$ pixels (matching our implementation) to obtain the filtered energy, $E_{\text{filtered}}$. The HF ratio is defined as
$\frac{E_{\text{filtered}}}{E_{\text{total}}}$, where higher values indicate that a larger fraction of the normal-map variation is concentrated in high-frequency components.

\myparagraph{Tiling.}
We evaluate tiling seamlessness using two metrics computed symmetrically along both horizontal and vertical seam boundaries. For a given texture image $I$, we analyze the transitions between opposing edges: $I_{\text{left}}$ vs. $I_{\text{right}}$ (horizontal) and $I_{\text{top}}$ vs. $I_{\text{bottom}}$ (vertical). The final score for each metric is the average of the horizontal and vertical measurements. Border continuity measures the pixel-level intensity jump directly at the seam interface. We compute the mean absolute difference between the pixels at the opposing boundaries for both directions and report the average: $$\mathcal{L}_{\text{border}} = \frac{1}{2} \left( \text{mean}(|I_{\text{left}} - I_{\text{right}}|) + \text{mean}(|I_{\text{top}} - I_{\text{bottom}}|) \right). $$ 
Gradient continuity quantifies how smoothly the local slope transitions across the seam when the texture is tiled. Beyond matching border pixels, a seamless tile should preserve the \emph{first-order difference} (local slope) across the wrap boundary. For left/right (x-direction) seams, we compare the gradient across the wrapped seam, $(I_{\text{left}}-I_{\text{right}})$, to the one-pixel internal gradients immediately inside each border, $(I_{\text{right}}-I_{\text{right}-1})$ and $(I_{\text{left}+1}-I_{\text{left}})$, and penalize their mismatch:
\begin{equation}
\begin{aligned}
\mathcal{L}_{\text{grad}}^{x}
= \mathrm{mean}\Big(
&\big| (I_{\text{left}}-I_{\text{right}}) - (I_{\text{right}}-I_{\text{right}-1}) \big| \\
+&\big| (I_{\text{left}+1}-I_{\text{left}}) - (I_{\text{left}}-I_{\text{right}}) \big|
\Big),
\end{aligned}
\end{equation}
where $I_{\text{right}-1}$ denotes the column one pixel inside from the right edge (and $I_{\text{left}+1}$ is one pixel inside from the left edge). The y-direction term $\mathcal{L}_{\text{grad}}^{y}$ is defined analogously for top/bottom seams. The final gradient continuity score is the symmetric average:
\[
\mathcal{L}_{\text{grad}}=\tfrac{1}{2}\big(\mathcal{L}_{\text{grad}}^{x}+\mathcal{L}_{\text{grad}}^{y}\big),
\]
where \(\mathrm{mean}(\cdot)\) averages over all seam pixels (and color channels). Lower values indicate fewer visible ``kinks'' at tile boundaries.

\begin{table}[t]
\centering
\setlength{\tabcolsep}{5pt}
\caption{\textbf{Tileability comparison of generated normal-map patches, including Qwen-Image-based Tiled Diffusion.} All baselines are initialized from the high-pass-filtered result.}
\label{tbl:tiling_metrics_supp}
\begin{tabular}{lccc}
\toprule
Method & Border Cont. $\downarrow$ & Grad Cont. $\downarrow$ & Run Time (s) \\
\midrule
Tiled Diffusion (SDXL)     & \underline{4.43}  & \underline{9.76}  & \underline{10.9} \\
Tiled Diffusion (Qwen-Image) & 22.02 & 44.83 & 93.71 \\
Ours (Inter-tile tiling)     & 7.19  & 14.42 & 24.88 \\
Ours (Intra-tile tiling)      & \textbf{1.61}  & \textbf{3.19}  & \textbf{3.18} \\
\bottomrule
\end{tabular}
\end{table}

In addition, we present results for Tiled Diffusion when reimplemented with Qwen-Image in~\reftbl{tiling_metrics_supp}, alongside the previously introduced SDXL (official) implementation. The core mechanism of Tiled Diffusion operates by extending the latent with padding on each side and copying edge strips to the opposite boundary padding at every denoising step, effectively enforcing the U-Net/DiT to always see continuous wrapped context at each border, similar in spirit to our inter-tile arrangement. In its img2img mode, it additionally employs Differential Diffusion~\cite{levin2025differential} to spatially modulate the denoising strength, which is essentially a soft mask that controls how much each region can change.

We hypothesize that our intra-tile arrangement outperforms both Tiled Diffusion and our inter-tile variant because the latter methods generate new latent values outside the image boundaries. Although the wrapping constraint encourages similarity, each opposing edge is still modeled as separate pixels and can diverge in shape and content. In contrast, the intra-tile arrangement generates no additional pixels. The borders are reorganized within the image through quadrant swapping, forcing each denoising step to directly produce the final output image.
We also observed that Differential Diffusion’s soft masking is less stable than hard masking and can occasionally introduce subtle artifacts. In addition, several components of Tiled Diffusion are tuned for the SDXL architecture and text-to-image generation task which has fewer constraints. Directly adapting them to Qwen-Image proved difficult because the two models differ substantially including the scheduler (DDPM-style vs. flow matching) and the backbone (U-Net vs. DiT). Despite extensive tuning, simple parameter adjustments did not lead to reliable improvements. For these reasons, the intra-tile arrangement combined with our masked inpainting and SDEdit formulation provides a simpler and more effective solution.

\subsection{Failure Cases of Other 3D Baselines}
\reffig{baseline_3d_failures} illustrates common failure modes of object-centric 3D generation models when adapted to 2.5D tactile representations. 
Although Hunyuan3D~2.1 produces visually plausible front views, side views reveal inconsistent or misplaced base plates that compromise physical stability. 
Trellis~2 is unable to generate valid geometry when a base plate is included in the input; even after manual adaptation to enforce a 2.5D structure, its outputs contain thin, fragmented, and non-manifold components that fail fabrication. 
These results highlight the challenges of directly applying general-purpose 3D generation models to fabrication-constrained tactile graphics and motivate the need for an explicitly 2.5D, fabrication-aware pipeline.
\begin{figure}[ht]
    \centering
    \includegraphics[width=0.7\textwidth]{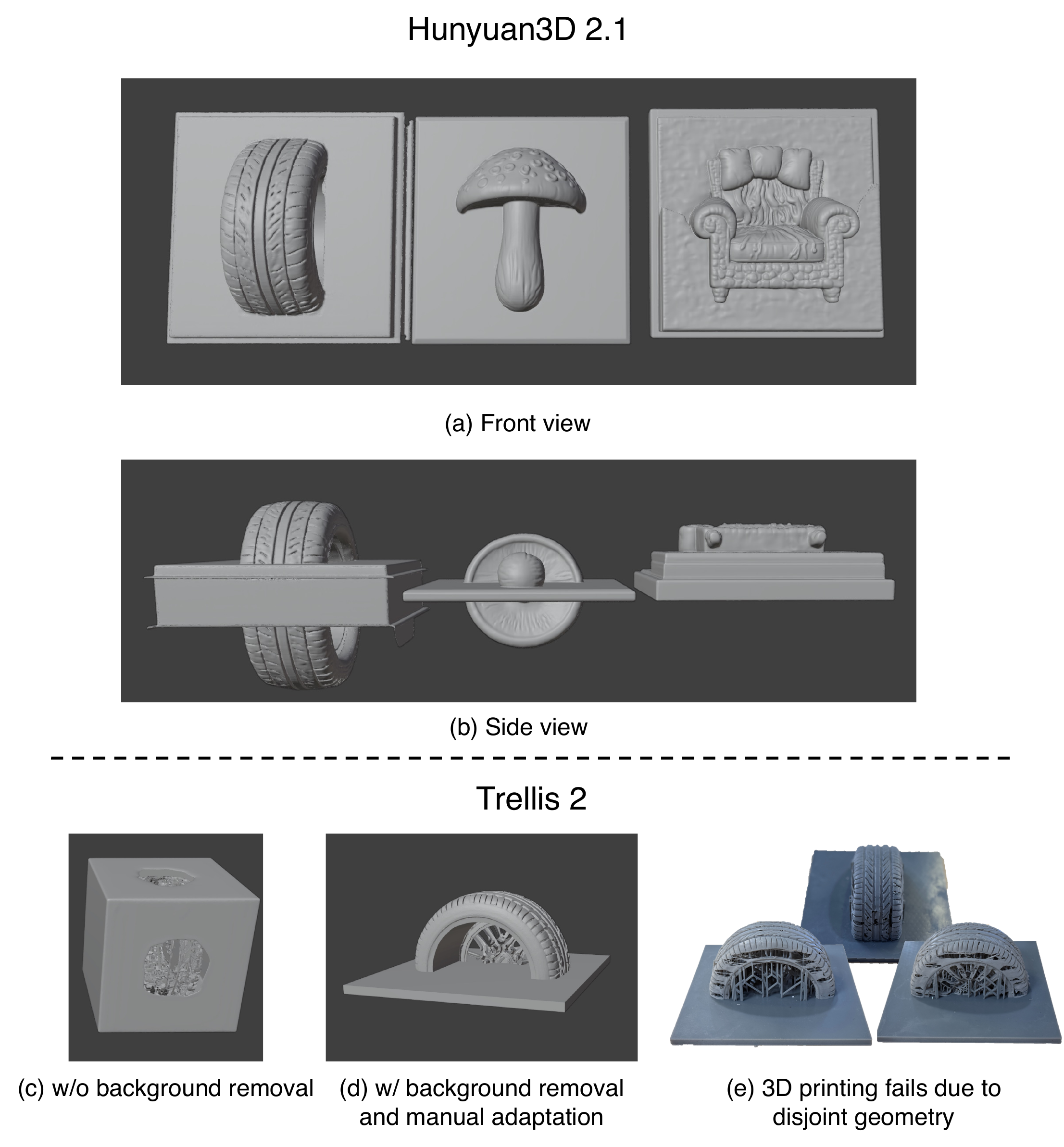}
    \caption{{\bf 3D baselines fail to produce stable 2.5D tactile representations.} 
    (a–b) Front and side views of results from Hunyuan3D~2.1. While the front views appear plausible, the side views reveal inconsistent placement of the base plate, leading to physical instability and extraneous bulk.
    (c–d) Results from Trellis~2 before and after manual adaptation. Trellis~2 fails when a base plate is included in the input image; to enable evaluation, we manually remove the background, generate relief geometry, and place it on a fixed base plate.
    (e) A representative 3D-printed Trellis~2 result, which fails fabrication due to thin structures and disjoint geometry. As a result, Trellis~2 is excluded from the in-person user study and used only for qualitative rendering comparisons. 
    }
    \lblfig{baseline_3d_failures}
\end{figure}

\section{Extended User Study Analysis}
\lblsec{extended_user_study}
Beyond quantitative ratings, the user study yielded detailed qualitative feedback that reveals how participants interpret and evaluate tactile textures in practice. Participants’ comments highlight not only perceptual preferences, but also how tactile graphics support imagination, material understanding, and exploratory engagement beyond everyday tactile experiences.

\subsection{Role of Braille in Semantic Understanding}
While object recognition accuracy is low without braille, BLV participants consistently reported that braille annotations fundamentally changed how the tactile graphics were interpreted.
Several participants noted that the object geometry became meaningful only after reading the label, describing the experience as the object ``making sense'' once semantic context was provided.
Participants also commented that braille enabled independent and self-paced exploration, without reliance on sighted assistance or audio descriptions.

\subsection{Sensitivity to Relief Amplitude}
Participants were sensitive to both insufficient and excessive surface relief.
Baseline methods were frequently described as overly smooth, making it difficult to perceive meaningful texture differences.
At the same time, participants occasionally noted that overly deep relief felt unnatural or distracting.
These observations suggest that tactile texture quality depends on a balanced relief amplitude rather than maximal geometric exaggeration.

\subsection{Role of Directionality and Anisotropy}
Textures with strong directional cues—such as longitudinal fibers, vertical ridges, or twisted patterns—were repeatedly identified as easier to interpret and more consistent with participants’ mental models of the target materials.
This indicates that preserving anisotropic structure in tactile textures plays a key role in perceptual clarity.

\subsection{Part-Aware Texture Comparison}
Participants often evaluated textures in relation to other parts of the same object, comparing, for example, the seat and frame of a chair or the cap and stalk of a mushroom.
Differences in texture structure across parts helped participants distinguish object regions and assess whether each surface matched its intended material, supporting the use of explicit, part-aware texture assignment.

\subsection{Imaginative and Inaccessible Tactile Experiences}
Participants distinguished between textures that corresponded to familiar real-world objects (e.g., football-like textures on a ball) and those representing materials that are inaccessible or nonexistent in everyday experience.
Examples included textures inspired by jellyfish tentacles, which are difficult or unsafe to touch in the real world, as well as imaginative combinations such as avocado-skin textures applied to ``dolphin with wings''.
Participants commented that interacting with these 3D-printed tactile graphics helped them imagine surfaces they had never touched before, describing the experience as engaging and enriching.
These observations suggest that generative tactile graphics can support not only realistic material representation, but also imaginative exploration beyond physical-world constraints.

\section{Additional Qualitative Results}
\lblsec{supp_additional_comparison_25D}
\subsection{Additional qualitative ablations of the base plate template}
\begin{figure}[t]
    \centering
    \includegraphics[width=0.7\textwidth]{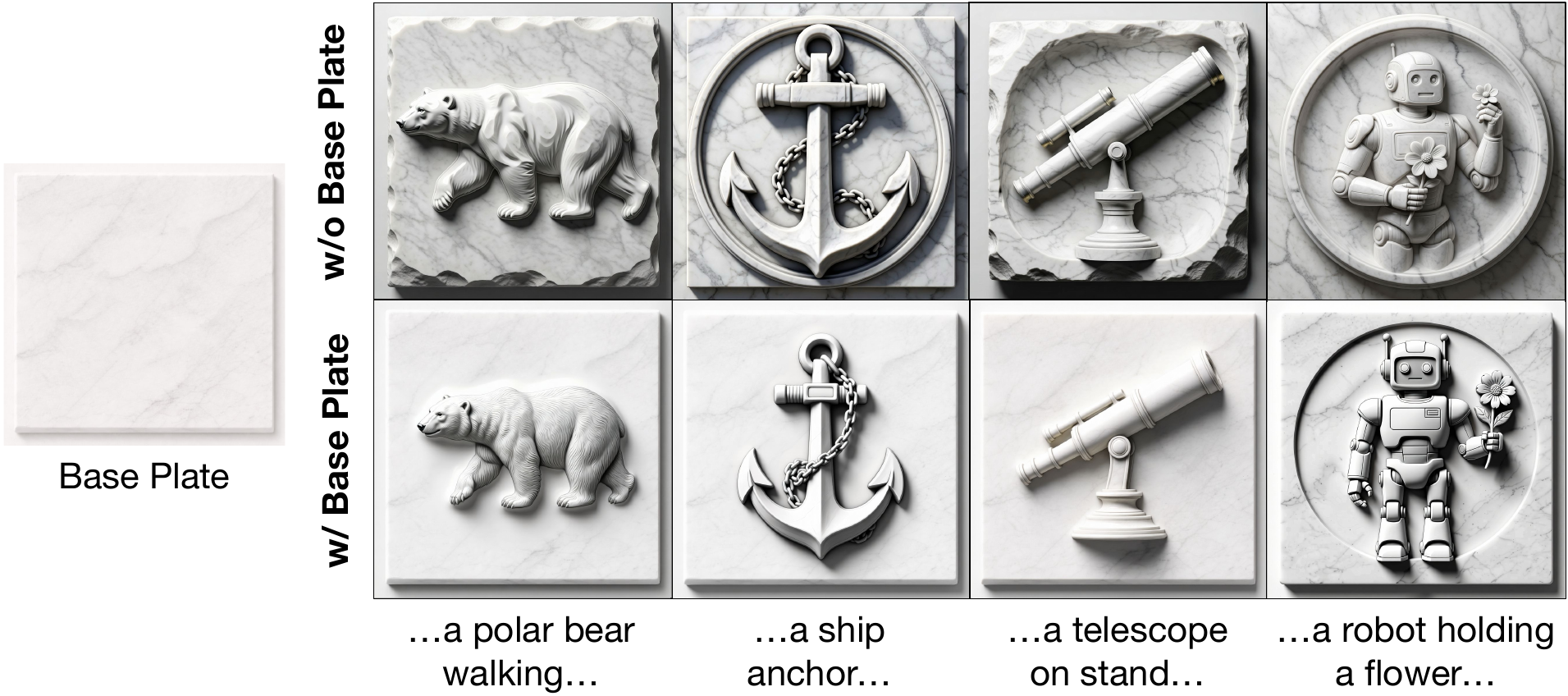}
    \caption{{\bf Ablation of the base plate template with Qwen-Image-Edit 40-step model.} Plate conditioning significantly reduces variation in plate height and improves geometric consistency compared to unconditioned generation.
    }
    \Description{Without conditioning on a base plate image, Qwen generates outputs with unnecessary background details, such as a carved edge around the main subject.}
    \lblfig{t2i_base_plate_qwen40}
\end{figure}

\begin{figure}[ht]
    \centering
    \includegraphics[width=0.7\textwidth]{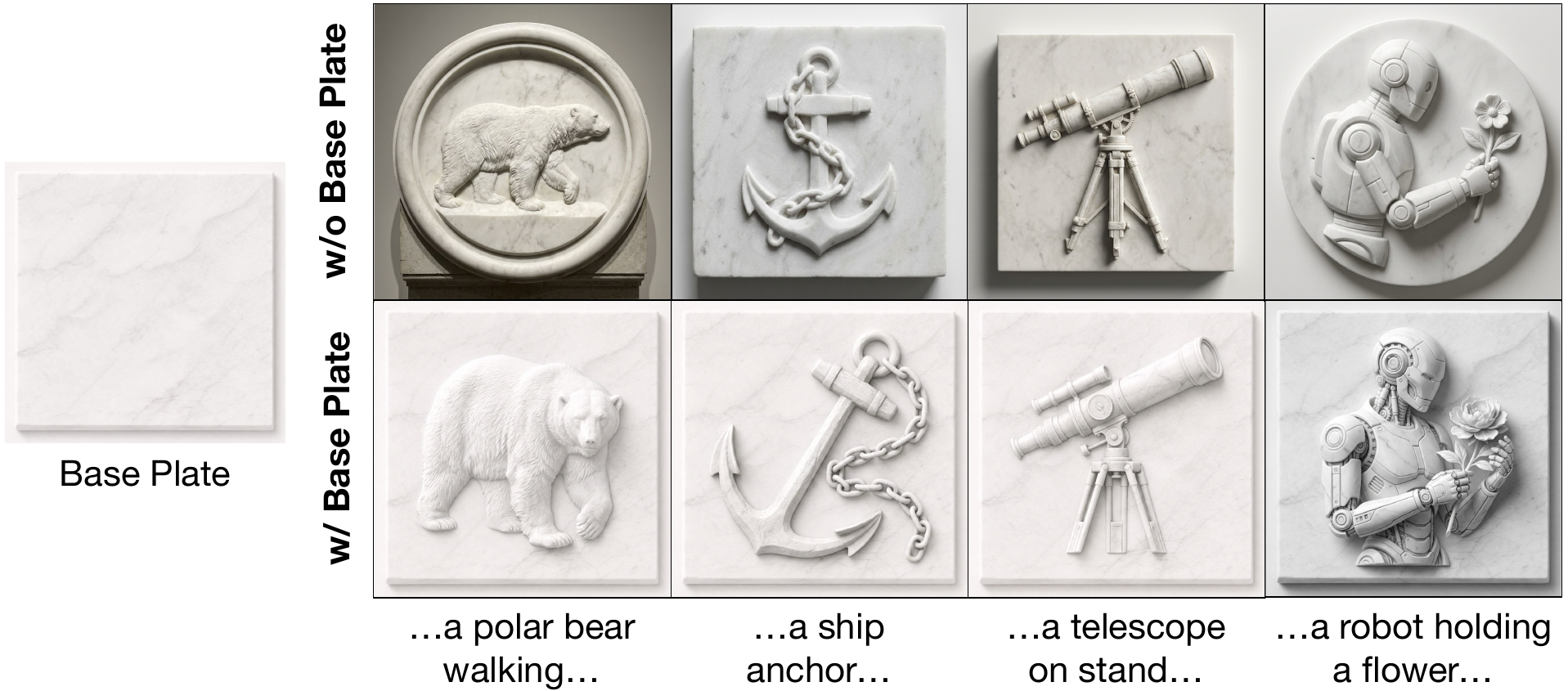}
    \caption{{\bf Ablation of the base plate template with Nano Banana Pro.} Enforcing a base plate prior yields stable plate geometry and consistent relief structure across diverse prompts.}
    \Description{Without conditioning on a base plate image, Nano Banana Pro generates outputs with inconsistent backgrounds unsultable for physical fabrication, such as placing the depicted sculpture on a stand.}
    \lblfig{t2i_base_plate_nb}
\end{figure}

\reffig{t2i_base_plate_qwen40} and \reffig{t2i_base_plate_nb} show qualitative ablations of the base plate template across two different text-to-image backbones. Without plate conditioning, generated reliefs exhibit large variations in plate height, curvature, and background geometry, leading to unstable or inconsistent 2.5D structures. 
Enforcing a base plate prior consistently produces flatter, well-aligned plates and more uniform relief geometry across diverse prompts, improving both geometric consistency and suitability for downstream fabrication.

\subsection{Comparison against traditional 2.5D representations}
While our system utilizes high-resolution 3D printing to fabricate 2.5D tactile graphics, we also evaluate its performance against traditional fabrication methods, specifically swell-paper printing. 
As discussed in \refsec{related_work}, swell paper is a widely adopted medium for creating affordable and durable tactile graphics. By running paper treated with black ink or toner through a fuser machine, the inked regions physically expand to create raised lines and diagrams.

We compare our method against TactileNet \cite{khan2025tactilenet}, a recent learning-based approach that fine-tunes Stable Diffusion using LoRAs and DreamBooth adapters for class-specific tactile image generation. Using their released weights, we generated two outputs per category:
\begin{itemize}
\item \textbf{Textureless/Raw prompts:} These describe only the global object shape, mimicking the original training distribution of TactileNet.
\item \textbf{Textured prompts:} These follow our proposed template by incorporating detailed material and texture descriptions.
\end{itemize}
The complete prompt pairs used for this comparison are detailed in \reftbl{TactileNet_prompt_pairs}.
\begin{table}[ht]
\centering
\small
\caption{Example prompt pairs for raw and textured tactile graphic generation.}
\lbltbl{TactileNet_prompt_pairs}
\begin{tabularx}{\textwidth}{l l X}
\toprule
\textbf{Category} & \textbf{Type} & \textbf{Prompt Description} \\ \midrule
\multirow{4}{*}{Chair} & Raw & Create a tactile graphic of a chair for visually impaired users, emphasizing the seat, backrest, and four legs, set against a plain background for clear contrast, raised smooth lines, simple silhouette. \\ \cmidrule{2-3}
 & Textured & Create a tactile graphic of a chair, the seat has fine cross-hatched raised lines, the legs have wicker basket texture, subtle vertical raised grain, the backrest has leather texture, evenly spaced horizontal raised slats, set against a plain background, bold outlines. \\ \midrule
\multirow{4}{*}{Lamp} & Raw & Create a tactile graphic of a lamp for visually impaired users, emphasizing the rounded base, vertical stem, and wide conical shade, set against a plain background, raised smooth lines, simple silhouette. \\ \cmidrule{2-3}
 & Textured & Create a tactile graphic of a lamp, the base has tree bark texture, vertical irregular raised ridges with rough fissured grooves, the shade has a cloth bag texture, loose woven grid of raised crossed lines, set against a plain background, bold outlines. \\ \midrule
\multirow{4}{*}{Canoe} & Raw & Create a tactile graphic of a canoe for visually impaired users, emphasizing the elongated hull, pointed bow and stern, and a paddle, set against a plain background, raised smooth lines, simple silhouette. \\ \cmidrule{2-3}
 & Textured & Create a tactile graphic of a canoe, the hull surface has an embossed flower texture, repeating pattern of raised petal outlines arranged in circular rosette clusters, set against a plain background, bold outlines. \\ \bottomrule
\end{tabularx}
\end{table}

\myparagraph{Evaluations and Results}
Tactile images generated by TactileNet were physically fabricated on swell paper using a Swell-Form machine.
\reffig{swell_paper_print} illustrates the results, including the generated digital images (a), the resulting swell-paper prints (b), a zoom-in view of the surface details (c), and the user study setup for BLV participants (d).
While images generated from textureless prompts (Fig. \ref{fig:swell_paper_print}a, left) yield reasonable silhouettes with clear outlines, TactileNet fails to generalize to textured prompts. 
In these cases, the model either misapplies the texture to the background (e.g., the floor in Rows 1 and 2) or fails to generate the primary object entirely (e.g., the canoe in Row 3).
\begin{figure}[t]
    \centering
    \includegraphics[width=0.99\textwidth]{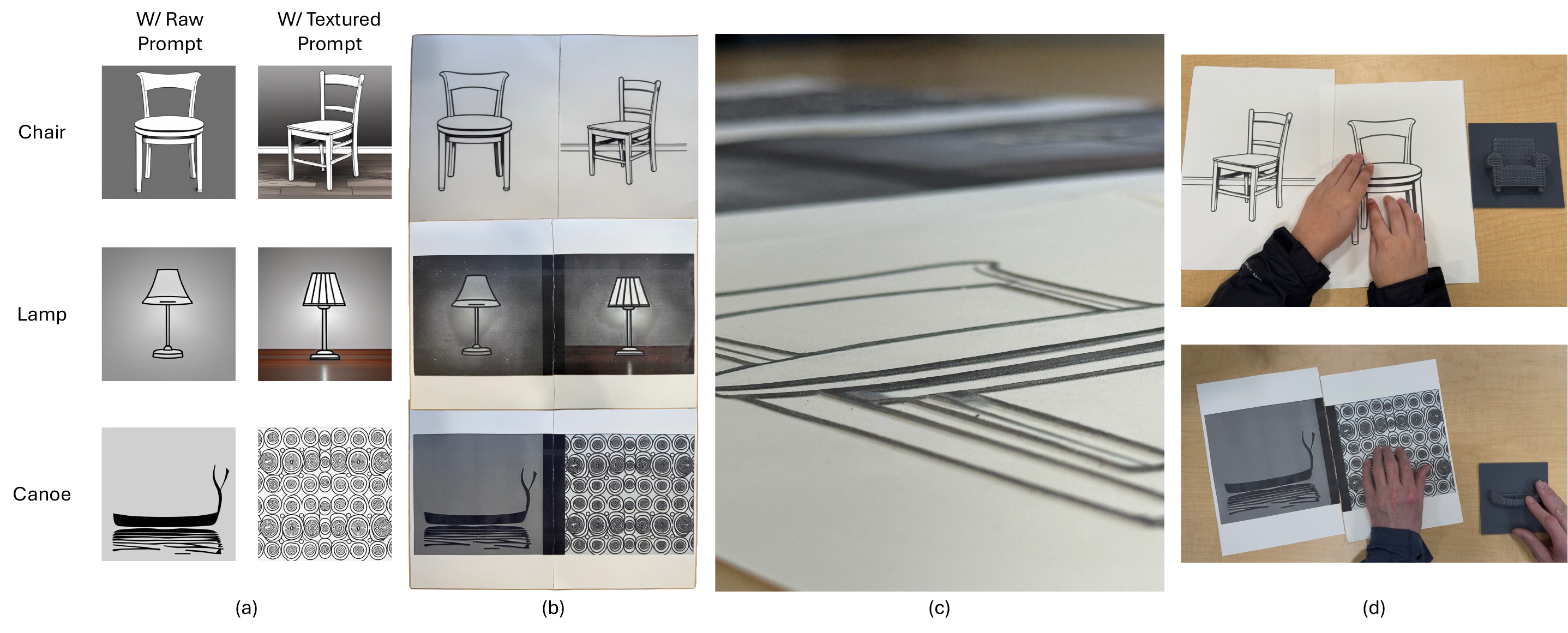}
    \caption{{\bf Comparison of generative tactile graphics on swell paper.} (a) Digital outputs generated by TactileNet~\cite{khan2025tactilenet} using class-specific LoRAs with ``textureless/raw'' (left) and ``textured'' prompts (right). While the model produces clear silhouettes for simple shapes, it fails to meaningfully ground complex texture descriptions, often misplacing them in the background. (b) Corresponding physical tactile graphics fabricated using a Swell-Form machine. For the chair (Row 1), backgrounds were manually removed to ensure printability. (c) Zoom-in view of the swell-paper surface, illustrating the raised lines created by the heat-reactive black toner. (d) User study setup where BLV participants evaluate the haptic recognizability and texture-matching of the fabricated samples. The detailed input text prompts are listed in \reftbl{TactileNet_prompt_pairs}.}
    \Description{Examples of swell paper printing using TactileNet.}
    \lblfig{swell_paper_print}
\end{figure}

To further evaluate these representations, we conducted a user study with two blind and low-vision (BLV) participants. 
Each user explored three formats of the same object: (1) swell-paper prints from textureless prompts (TactileNet), (2) swell-paper prints from textured prompts, and (3) our 3D-printed 2.5D graphics from textured prompts. 
Participants rated each print on a 5-point Likert scale based on how effectively the physical output matched the intended shape and texture. 
Detailed results are provided in \reftbl{user_study_swell_paper_result}.

In qualitative interviews, participants noted that our 3D-printed representation provided a significantly more enriched haptic experience. 
Specifically, the 3D prints allowed for more diverse patterns, such as the irregular vertical ridges of the tree bark texture, which were easier to differentiate than the limited relief of swell paper. 
Furthermore, participants found it easier to recognize complex structures—such as the distinct parts of a chair—because 3D printing supports refined depth variations and smoother transitions. 
This enabled a more holistic ``full-hand'' exploration of global geometry, whereas swell paper often restricted users to purely contour-based tracing.

\begin{table}[t]
\centering
\setlength{\tabcolsep}{5pt}
\caption{\textbf{Perceptual Study on 2.5D VS 3D representation.}}
\lbltbl{user_study_swell_paper_result}
\begin{tabular}{lccc}
\toprule
Attribute & 2.5D TactileNet & 2.5D Textured Prompt & 3D (Ours) \\
\midrule
Geometry  & $2.00 \pm 0.89 $ & $1.83 \pm 0.98 $ & $3.50 \pm 0.55 $ \\
Texture   & $1.67 \pm 0.82 $ & $2.67 \pm 1.37 $ & $3.50 \pm 1.38 $ \\
\bottomrule
\end{tabular}
\end{table}

\subsection{Comparison against Expert-Designed Swell Paper Graphics}
\lblsec{artist_swell_comparison}
While the preceding comparison evaluates against learning-based swell paper generation (TactileNet), we additionally compare our 3D-printed tactile graphics against \emph{expert-designed} swell-paper artifacts to assess performance relative to the current gold standard in manual tactile graphic production.

\myparagraph{Stimulus Design.}
For each of five objects (\emph{Lamp, Canoe, Football, Polar bear, Coral tree}), we prepared two swell-paper variants:
(1)~\textit{Swell (textureless)}: a line drawing sourced from the APH Tactile Graphic Image Library, an established repository of expert-curated tactile graphics for the BLV community;
(2)~\textit{Swell (textured)}: the same drawing with material textures manually applied to the corresponding regions by a sighted artist, using the same texture descriptions as our generated artifacts.
Both variants were physically fabricated on a Swell-Form machine, following the same protocol as the TactileNet comparison above.

\myparagraph{Procedure.}
Two BLV participants each evaluated all three conditions (textureless swell, textured swell, and our 3D-printed graphic without braille) on the five objects, yielding $N{=}10$ ratings per condition.
Participants rated each artifact on a $1$--$5$ Likert scale along three dimensions: \emph{object recognition} (how well the shape conveys the depicted object), \emph{geometry/shape match} (fidelity of the overall form), and \emph{texture realism} (how well the surface material matches the intended texture).

\myparagraph{Results.}
As shown in \reftbl{artist_swell_comparison}, our method is rated highest across all three attributes.
On \emph{geometry}, our 3D prints score $4.80$ versus $4.30$/$4.10$ for the two swell-paper variants. The relatively small gap is expected, as expert-designed line drawings produce geometrically faithful silhouettes.
On \emph{texture realism}, the textureless swell variant scores poorly ($3.10$) because it carries no material information, while the artist-textured variant benefits from manual texture application ($4.80$). Our 3D prints attain a perfect $5.00 \pm 0.00$, demonstrating that the high-resolution SLA printing effectively conveys fine surface detail.
On \emph{recognition without braille}, our method also scores highest ($4.50$ versus $3.70$/$3.80$), suggesting that the richer tactile cues from 3D printing support better object identification even in the absence of semantic labels.

\begin{table}[t]
\centering
\setlength{\tabcolsep}{5pt}
\caption{\textbf{Perceptual study comparing our 3D-printed graphics against artist-designed swell-paper graphics.} Likert scale $1$--$5$ (higher is better); $N{=}10$ ratings per cell (2 BLV participants $\times$ 5 objects). \textit{Swell (textureless)} uses expert-curated line drawings from the APH Tactile Graphic Image Library; \textit{Swell (textured)} applies material textures to the same drawings by a sighted artist.}
\lbltbl{artist_swell_comparison}
\begin{tabular}{lccc}
\toprule
Attribute & Swell (textureless) & Swell (textured) & 3D (Ours w/o Braille) \\
\midrule
Recognition & $3.70 \pm 0.67$ & $3.80 \pm 1.03$ & $\mathbf{4.50 \pm 0.85}$ \\
Geometry    & $4.10 \pm 0.74$ & $4.30 \pm 0.95$ & $\mathbf{4.80 \pm 0.63}$ \\
Texture     & $3.10 \pm 0.88$ & $4.80 \pm 0.42$ & $\mathbf{5.00 \pm 0.00}$ \\
\bottomrule
\end{tabular}
\end{table}

In qualitative interviews, participants emphasized texture and spatial three-dimensionality as the primary differentiating factors. On the coral tree, one participant noted that the rough surface of the 3D print helped identify the object, while the swell-paper version only conveyed a generic plant outline. On the polar bear, a participant described the 3D-printed legs as having a distinct tactile quality absent from the flat line drawing. These observations indicate that our system's primary advantage over expert-designed swell paper lies not in shape fidelity, but in conveying surface materials beyond what flat heat-expanded ink can achieve.

\subsection{Comparison with Material- and Texture-Generation Baselines}
\lblsec{material_gen_comparison}

\begin{figure}[t]
    \centering
    \includegraphics[width=0.85\textwidth]{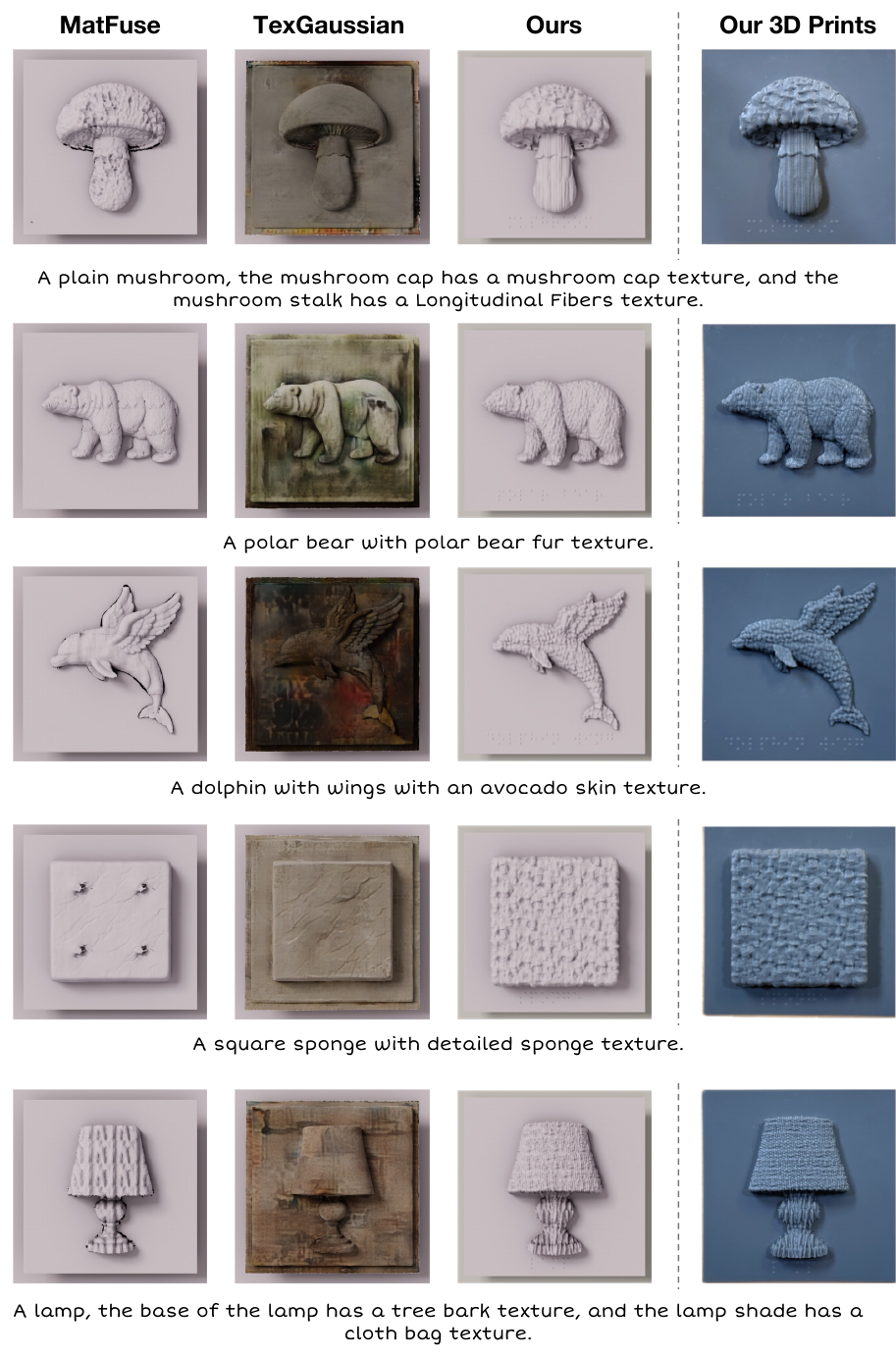}
    \caption{\textbf{Qualitative comparison with material/texture-generation baselines.} From left to right: \textbf{MatFuse}~\cite{vecchio2024matfuse} (2D SVBRDF, with its normal map applied to our base mesh), \textbf{TexGaussian}~\cite{xiong2025texgaussian} (text-conditioned 3D PBR painting), \textbf{Ours}, and a photograph of our 3D-printed prototype. Rows show representative objects (mushroom, polar bear, winged dolphin, sponge, and lamp). MatFuse and our results are shown with a neutral matte clay shader to expose surface geometry; TexGaussian is shown with its native albedo/roughness/metallic appearance, since it produces color rather than relief.}
    \Description{Qualitative comparisons against material generation baselines.}
    \lblfig{material_gen_comp}
\end{figure}

\myparagraph{MatFuse.}
MatFuse~\cite{vecchio2024matfuse} is a 2D SVBRDF diffusion model that given a text
prompt, generates a set of spatially-varying material maps (diffuse albedo,
tangent-space normal, roughness, and specular). 
We use the official checkpoint and condition it on the \emph{same} per-segment texture prompts used by our texture stage. 
Since MatFuse is purely image-space and synthesizes no geometry, we add a texture-application step so that it can be evaluated as a 3D tactile relief. 
We take MatFuse's tangent-space normal map, tile it across the segmented foreground region, and integrate it into a height field with the same normal-to-displacement integration our method uses. 
This displacement is then applied to \emph{our} base relief mesh with the same plate flattening (without braille) for a fair comparison. 
This grants MatFuse the benefit of our geometry pipeline and isolates the quality of its predicted surface micro-structure.

\myparagraph{TexGaussian.}
TexGaussian~\cite{xiong2025texgaussian} is a feed-forward 3D PBR texture generator that given an input mesh and a text prompt, predicts surface material maps (albedo, roughness, metallic) using an octree-based 3D Gaussian representation. 
We use the official checkpoint, provide it with our Stage-1 base mesh and object prompt used by all methods, and export the resulting textured mesh.
By construction, TexGaussian only \emph{paints} materials onto the supplied geometry and does not modify the surface relief.

\myparagraph{Rendering protocol.}
All meshes share one Blender~Cycles setup (camera, lighting, backdrop, and a front-facing view with normalized footprint). 
To isolate tactile geometry, our results and the MatFuse relief use a neutral matte clay shader, while TexGaussian is shown in its native PBR appearance.

\myparagraph{Results.}
The comparison shown in \reffig{material_gen_comp} highlights why color- and material-centric generators are ill-suited to tactile graphics. 
TexGaussian re-paints the base silhouette with a 2D appearance, but it fails to understand the shape of a 2.5D representation and its surface remains essentially flat, so the ``texture'' exists only as baked shading that a blind reader cannot perceive by touch. 
MatFuse does inject some relief through its normal map, but the recovered geometry is
low-frequency, soft, and frequently noisy or semantically incoherent, with broken contours and stray bumps, e.g., the sponge and dolphin rows.
It also fails to reproduce the distinct, finger-discriminable patterns the prompts call for. 
In contrast, our method produces crisp, high-frequency, and semantically aligned tactile relief, such as the woven lamp shade, the bear's fur, the sponge's pores, etc., which is both legible to the touch and faithfully preserved after fabrication, as confirmed by the 3D-printed prototypes.

\end{document}